\documentclass[]{raa}            
\usepackage{graphicx,times}
\usepackage{natbib}

\providecommand{\bm}[1]{\mbox{\boldmath$#1$\unboldmath}}

\begin{document}

   \title{Chemical mixing by turbulent
   convection in the overshooting region below the
   convective envelope of RGB stars\footnote{Supported by the National Natural Science Foundation of China.}
}

 \volnopage{ {\bf 2011} Vol.\ {\bf X} No. {\bf XX}, 000--000}
   \setcounter{page}{1}

   \author{X.-J. Lai
      \inst{1,2,3}
   \and Y. Li
      \inst{1,2}
   }

   \institute{National Astronomical Observatories/Yunnan Observatory, Chinese Academy of Sciences, Kunming 650011, China;
  {\it lxj@mail.ynao.ac.cn, ly@ynao.ac.cn}\\
        \and
             Key Laboratory for the Structure and Evolution of
             Celestial Objects, Chinese Academy of Sciences, Kunming 650011, China
        \and
             Graduate School of the Chinese Academy of Sciences, Beijing 100049,
             China.
\vs \no
   { }
}

\abstract{ Based on the turbulent convection model (TCM), we
investigate  chemical mixing in the bottom overshooting region of
the convective envelope of intermediate-mass stars, focusing on its
influence on the formation and extension of blue loops in the
Hertzsprung-Russell (HR) diagram. A diffusive mixing model is
adopted during the Red Giant Branch (RGB) phase. The properties of
the blue loop are changed by modification of the element profiles
above the H-burning shell, which results from the incomplete mixing
in the bottom overshooting region when stellar model evolves up
along the RGB. Such modification of the element profiles will lead
to an increase of the opacity in the region just above the H-burning
shell and a decrease of the opacity in the outer homogeneous
convection zone, which will result in a quick decrease of the
H-shell nuclear luminosity $L_{H}$ when the stellar model evolves
from the RGB tip to its bottom and, finally, a much weaker and
smaller convection zone will be obtained in the stellar envelope.
This helps to form a longer blue loop. The extension of the blue
loop is very sensitive to the parameters $(C_{X}$ and
$\alpha_{TCM})$ of the diffusive mixing model and of the TCM. The
results mainly show that: 1) comparing to the results of the
classical model with the mixing-length theory, the lengths of the
obtained blue loops with different combinations of the values of
$C_{X}$ and $\alpha_{TCM}$ are all increased and the length of the
blue loop increases with the values of parameters
 $C_{X}$ and $\alpha_{TCM}$; 2)
the diffusive mixing model can significantly extend the time of
stellar models lingering on the blue side of the HR diagram, even
though the length of the blue loop for the 7$M_{\odot}$ star has a
less prominent difference between the classical and diffusive mixing
model; 3) both the observations referring to the location of the
Cepheid instability strip and the number ratio $N_{B}/N_{R}$
 of blue to red evolved stars in the Galactic open clusters can
 confine the two parameters in a range of $0.5\leq\alpha_{TCM}\leq0.9$
 and $10^{-5}\leq C_{X}\leq10^{-4}$ for the model of $5M_{\odot}$. However, for the case of the $7M_{\odot}$ star, there seems to be no such a definite range to even only account for the observed number
 ratio $N_{B}/N_{R}$. In any case, our results based on the diffusive mixing model are on the whole in accordance with not only other theoretical ones but also the observations.
  }

   \authorrunning{X.-J. Lai \& Y. Li }            
   \titlerunning{Chemical mixing by turbulent convection in the overshooting region }  
   \maketitle


%
%
\section{Introduction}           
\label{sect:intro} Convection in stellar models is an important
mechanism of element mixing. The mixing-length theory (MLT,
B\"{o}hm-Vitense 1958) is commonly adopted to depict the turbulent
convection process in the stellar evolution models. The boundaries
of a convection zone are defined by a local stability criterion,
i.e. the Schwarzschild condition $\nabla_{rad}=\nabla_{ad}$ where
$\nabla_{rad}$ and $\nabla_{ad}$ are the radiative and adiabatic
temperature gradient, respectively. In such way the interface
between a radiative and a convection region is sharply defined.
However, in reality because of the dynamical consequence of Newton's
law (Canuto 1998) some convective elements may penetrate into the
radiative equilibrium region, resulting in a so-called 'overshooting
region' between the unstably convective zone and the stably
radiative one, which is widely validated in real stars by
comparisons of the stellar models with observations (Maeder \&
Meynet 1989; Stothers 1991; Alongi et al. 1991; Schr\"{o}der et al.
1997; Herwig 2000). Both the thermal structure and the chemical
composition will be modified due to the overshooting effects, and
thus the stellar structure and evolution will be significantly
affected. However, there is no widely accepted theoretical model to
reliably describe the convective overshooting nowadays. The focused
debates are the extent of the overshooting and the efficiency of the
mixing induced by it (Saslaw \&  Schwarzschild 1965; Shaviv \&
Salpeter 1973; Eggleton 1983; Renzini 1987; Xiong 1985,1986; Deng et
al. 1996a, 1996b; Freytag et al. 1996; Canuto 1998; Salasnich et al.
1999; Robinson et al. 2004). Because of the lack of a fully sounded
theory, some parameterized approaches have to be adopted, in most of
which the local pressure scale height is usually taken as a
parameter to describe the overshooting distance. It is needed to
stress here that the problem of the convective overshooting is just
induced by some weakpoints or inherent inconsistencies of the MLT
itself (Deng et al. 1996a), which can be ruled out by the turbulent
convection theory.

According to the MLT, chemical elements are mixed instantaneously
and homogeneously in a convection zone (Vandenberg \& Smith 1988;
Sweigart 1997). However, this assumption should not be adopted
straightforward for the overshooting regions (Deng et al. 1996a).
Because both laboratory experiments and the numerical simulations
suggest that the speed of mixing is slow enough that instantaneous
mixing does not achieve. Therefore, mixing in the overshooting
regions are often approximated by a time-dependent diffusion
process. However, the diffusion coefficient is often assumed as
consisting of an exponential decay of convective velocity within the
region (Freytag et al. 1996; Herwig 1997,2000; Salasnich et al.
1999; Ventura \& D'Antona 2005). Deng et al. (1996a,b) proposed an
improved formalism for the diffusion coefficient, which includes the
contribution of the motions of all possible scales and the effects
of intermittence and stirring of turbulence.

Different ways of convective mixing (instantaneous or diffusive) and
different values of the diffusion coefficient for the diffusive
mixing may result in significant differences in stellar structure
and evolution, for example, the predicted location of the
instability strip for Cepheids, the extent of blue loops in the HR
diagram and the predicted number ratio of stars in the blue and red
part of the HR diagram. For the Galactic classical Cepheids,
different kinds of their masses are obtained by Caputo et al.
(2005), such as the pulsation mass, the evolution mass, from which
some low masses down to 5M$_{\odot}$ are got, or even lower ones.
Nevertheless, standard stellar models (e.g. no overshooting, no mass
loss, etc) fail to reproduce those low mass Galactic classical
Cepheids (Schmidt 1984; Fernie 1990). Because the blue tips of the
blue loops of the evolutionary tracks for these less massive models
can not go hot enough to approach the blue edge of the observed
Cepheid instability strip. And the smaller a model mass, the shorter
the length of its blue loop, even no blue loop develops. However,
the blue loop can occur even down to 4M$_{\odot}$ from the
observational data (Stothers \& Chin 1993). Many input physical
factors can affect the blue loop and its extension, such as initial
chemical composition, rates of $3\alpha$,
$C^{12}(\alpha,\gamma)O^{16}$ and $N^{14}(p,\gamma)O^{15}$
reactions, envelope opacities, mass loss and rotation. The detailed
discussions can be found in Stothers \& Chin (1994), El Eid (1995),
Xu \& Li (2004a,b) and Valle et al. (2009). It is necessary to point
out that the effects of the above mentioned factors on properties of
blue loop are still matters of debate. For example, for stars in the
core He burning phase, some authors (e.g. Brunish \& Becker 1990; Xu
\& Li 2004a) found that the blue loop is very sensitive to the
$C^{12}(\alpha,\gamma)O^{16}$ reaction rate and the length of the
blue loop increases with it. However, Valle et al. (2009) recently
showed that according to the updated rate of
$C^{12}(\alpha,\gamma)O^{16}$ from Hammer et al. (2005) the effect
of the reaction on the blue loop length is negligible for stars of
all masses. Convective mixing is a special one affecting the blue
loop, including convective criterion (Schwarzschild or Ledoux),
seimconvection, outward overshooting from the convective core and
downward overshooting from the convective envelope. In the case of
intermediate-mass stars the lengths of blue loops are found to be
longer by using Schwarzschild criterion than the ones by using
Ledoux criterion, whereas the reverse results are found in massive
stars (El Eid 1995). This is because that in massive stars an
intermediate convective zone is formed by using Schwarzschild
criterion which inhibits the downward penetration of the convective
envelope and suppresses the blue loop (Ritossa 1996). For both
intermediate and massive stars the extension of the blue loop will
be increased by adding an additional efficient semiconvection mixing
to the Ledoux convection (El Eid 1995; Deng et al. 1996a). The
effect of overshooting from the convective core during the main
sequence phase suppresses the blue loops (Matraka et al. 1982; Huang
\& Weigert 1983; El Eid 1995; Valle et al. 2009). However, The
overshooting below the convective envelope favors the development of
the blue loop (Stothers \& Chin 1991; Alongi et al. 1991) even in
models with core overshooting (Alongi et al. 1991; Deng et al.
1996b). And Alongi et al. (1991) found this factor to be the most
relevant one in producing the blue loop.

On the observational side, the number ratio $N_{B}/N_{R}$ of blue to
red giants is one of the observable stellar properties. It can be
taken as a probe of the stellar structure of intermediate-mass
stars, because it is very sensitive to the convective mixing (Langer
\& Maeder 1995; Deng et al. 1996b). Its value was first shown to
vary through galaxies by Vandenbergh (1988). Besides this, for a
given luminosity range its value increases with metallicity (Langer
\& Maeder 1995 for a review). Similarly for a given age range its
value increases with metallicity and decreases with distance from
galactocenter (Eggenberger et al. 2002). However, the value of
$N_{B}/N_{R}$  for a cluster suffer many uncertainties: i.e.
incompleteness of data, difference between photometric and
spectroscopic results, contamination of filed stars, dynamical
evolution of the cluster and so on. On the theoretical side, the
ratio $\tau_{He}^{B}/\tau_{He}^{R}$ of the time spent at high
effective temperature range to the time spent at low effective
temperature range during the core helium burning phase is expected
to be a close measure of the observational counterpart $N_{B}/N_{R}$
(Stothers 1991; El Eid 1995; Langer \& Maeder 1995). As the
convective mixing is a very sensitive factor to affect the extension
of the blue loop, it of course affects the value of
$\tau_{He}^{B}/\tau_{He}^{R}$.

It is the aim of the present paper to study the effect of the
chemical mixing in the outer convective envelope on the structure
and evolution of intermediate-mass stars by use of the TCM proposed
by Li \& Yang (2007). We propose a new diffusive mixing model based
on a new diffusion coefficient to describe the chemical mixing in
the convective overshooting region, while within the convective
envelope we simply use the homogeneous mixing model. Under such
considerations extended blue loops are obtained. And then we make
some comparisons between the obtained results and the related
observational results: location of the instability strip for the
Cepheids, and number ratio of stars in the blue and red part of the
HR diagram.

The plan of this paper is organized as follows. In Section 2 we
briefly describe the diffusion algorithm in the overshooting region.
The information of the evolutionary code and input physics and some
physical assumptions are described in Section 3. The evolutionary
results about the blue loop depending on the diffusion parameters
$C_{X}$ and $\alpha_{TCM}$ are given in Section 4 and discussed in
detail in Section 5. Comparisons with observations are presented in
Section 6. Finally, some concluding remarks are summarized in
Section 7.


\section{Chemical mixing by turbulent convection}
\label{sect:Obs}

The intermediate-mass stars develop large-scale convection in the
outer envelope during the RGB stage. The whole convective envelope
can be determined by the TCM without the aid of any instability
criterion, however, it is convenient to subdivide it into a
traditional convective zone (unstable stratification) and
overshooting region (stable stratification). In the convection zone
elements are assumed to be instantaneously homogenized due to
efficient mixing, while in the overshooting region we adopt a
diffusion equation to describe such incomplete mixing process:
\begin{equation}
\frac{\partial X}{\partial t}=\frac{\partial}{\partial
m_{r}}\biggl[(4\pi r^{2}\rho)^{2}D\frac{\partial X}{\partial
m_{r}}\biggl], \label{eq:LebsequeI}
\end{equation}
where $X$ is the mass fraction of the chemical element under
consideration and $D$ is the diffusion coefficient. Certainly the
parameter $D$ plays a crucial role in determining the extent of the
chemical mixing and it needs to be properly assigned according to
the TCM.

It needs to point out that Eq.(1) is only invoked to describe the
incomplete mixing in the overshooting region below the convective
envelope. We simply assume that the complete mixing extends to the
stellar surface. So the following discussion on the diffusion
coefficient $D$ is restricted only in the bottom overshooting
region. Hydro-dynamical simulations show that convective turbulent
velocity decays exponentially into the nearby stable radiative zone
(Freytag et al. 1996), so the formula of $D$ is usually constructed
in light of the behavior of the turbulent velocity in the
overshooting region. Different authors in fact adopt different
e-folding distances of the decay according to the pressure scale
height at the convective boundary (Freytag et al. 1996; Herwig et
al. 1997; Salasnich et al. 1999; Ventura \& D'Antona 2005). However,
all those diffusion coefficients use boundary values of convective
velocity according to the MLT.

In the present paper we propose a new formula of the diffusion
coefficient $D$ in the framework of the TCM. Because the convective
velocity in the overshooting region has been self-consistently set
up in advance, we do not need to artificially construct the decay of
the convective velocity, and furthermore the dimension of the
overshooting region $L_{OV}$ is correspondingly determined, which
covers the region from the border of the convective unstable region
to the location with zero convective velocity. Our diffusion
coefficient is
\begin{equation}
D=C_{X}\frac{\overline{u_{r}^{'}u_{r}^{'}}}{\sqrt{k}}l,
\label{eq:LebsequeI}
\end{equation}
where $C_{X}$ is an adjustable parameter that determines the
efficiency of convective mixing in the overshooting region,
$\overline{u_{r}^{'}u_{r}^{'}}$ is the radial turbulent kinetic
energy, $k$ is the turbulent kinetic energy and $l$ the typical
length of the turbulent mixing. Casting light on the mixing length
parameter $l$ there are two choices that can be used to determine
it. 1) $l=\alpha_{TCM}H_{P}$, where $\alpha_{TCM}$ is a parameter
introduced in the TCM and is similar to the parameter $\alpha$ in
the MLT, and $H_{P}$ is the local pressure scale height. 2)
$l=L_{OV}$, the length of the overshooting region that can be
obtained by the profile of the convective velocity in the
overshooting region. However, the bottom of the whole convective
envelope can numerically be set to the location, for example, where
$\sqrt{k}\sim 10^{-10}$cm/s. The advantage of using $l=L_{OV}$ is to
omit an adjustable parameter.

It should be noted that there are three advantages in our new
construction of the diffusion coefficient $D$ compared to other
mixing models mentioned above. First, the decaying way of $D$ owing
to the behavior of the convective velocity
$(\sqrt{u_{r}^{'}u_{r}^{'}}/\sqrt{k})$ is self-consistently obtained
by the TCM. Second, the dimension $(L_{OV})$ of the overshooting
region can be ascertained. Third, the value of convective velocity
at the boundary of the convective unstable zone is instead given by
the TCM.

\section{Evolutionary code and input physics}
\label{sect:data}

The stellar evolution code is originally described by B. Paczynski
and M. Kozlowski, and updated by R. Sienliewicz. Nuclear reaction
rates are adopted from BP95 (Bahcall, Pinsonneault \& Wasserburg
1995). The equation of state is the OPAL equation of state from
Rogers (1994) and Rogers, Swenson \& Iglesias (1996). The OPAL
opacities GN93hz series (Rogers \& Iglesias 1995; Iglesias \& Rogers
1996) are used in the high-temperature region. In the outer envelope
of the considered models, low-temperature opacities from Alexander
\& Ferguson (1994) are used. Some recent improvements are given by
Li \& Yang (2007), Zhang \& Li (2009). Here we list some details
connecting with the formation and development of the blue loop and
represent some input physics.

\hangafter=1\setlength{\hangindent}{2.8em}1) In order to reveal the
effect of the chemical mixing in the outer convective envelope on
the extension of the blue loop, we only apply the TCM in the
envelope convection, not in the convective core. And the convective
mixing begin to be incorporated into the evolutionary model when a
stellar model arrives at the bottom of the RGB in the HR diagram.
This is because that the outer convective envelope is very shallow
or does not even exist before this evolution phase location, and the
outer convective envelope can not approach the chemical gradient
region left by the central H burning stage. Therefore, the
evolutionary results will be completely the same for the stellar
models with or without the chemical mixing in the outer convective
envelope.

\hangafter=1\setlength{\hangindent}{2.8em}2) Because of some
numerical problems the TCM can not be used to predict the
temperature structure in our RGB models. Instead, the MLT is used in
our stellar evolutionary calculations, and the mixing-length
parameter $\alpha$ in the MLT is adopted to be 1.7. When each model
after the RGB bottom is obtained based on the MLT, the TCM is
applied in the convective envelope to determine the diffusion
coefficient $D$, and then the chemical structure is obtained to
complete one time step of the model's evolution. However, the
obtained results will have no significant changes under this
simplification because in such area the actual temperature gradient
is very close to the adiabatic one owing to very high efficiency of
convective energy transport during the RGB phase. We only follow
changes of $H$, $^{3}He$, $^{4}He$, $^{12}C$, $^{14}N$ and $^{16}O$
for the sake of simplicity.

\hangafter=1\setlength{\hangindent}{2.8em}3) The rates of $3\alpha$,
$C^{12}(\alpha,\gamma)O^{16}$ and $N^{14}(p,\gamma)O^{15}$ reactions
are taken from Caughlan \& Fowler (1988). The screening factors are
from Graboske et al. (1973). The neutrino emission rates are adopted
from Beaudet et al. (1967) and include the later correction factors
derived by Ramadurai (1976) for taking the neutral currents into
account.

\hangafter=1\setlength{\hangindent}{2.8em}4) Evolutionary models are
calculated for intermediate-mass stars of $4$, $5$ and $7M_{\odot}$
with (X, Z)=(0.28, 0.02). A set of the TCM's parameters are adopted:
$C_{t1},C_{e1},C_{s}=0.0313$, $C_{t}=3.0$, $C_{e}=1.25$,
$C_{k}=2.5$. Additionally, the parameter $C_{X}$ in the diffusive
mixing model and $\alpha_{TCM}$ in the TCM are given respectively:
$C_{X}=10^{-4}, 10^{-5}, 10^{-6}$; $\alpha_{TCM}=0.9, 0.5, 0.2$.
Therefore, there are $3\times 3$ different models for the evolution
series of a special stellar mass. Mass loss by the stellar wind is
not considered in this paper.
\section{Results of the stellar structure and evolution}
   \begin{figure}[h!!!]

    \begin{minipage}[t]{70mm}
    \centering

   \includegraphics[width=7.2cm, height=62mm,angle=0]{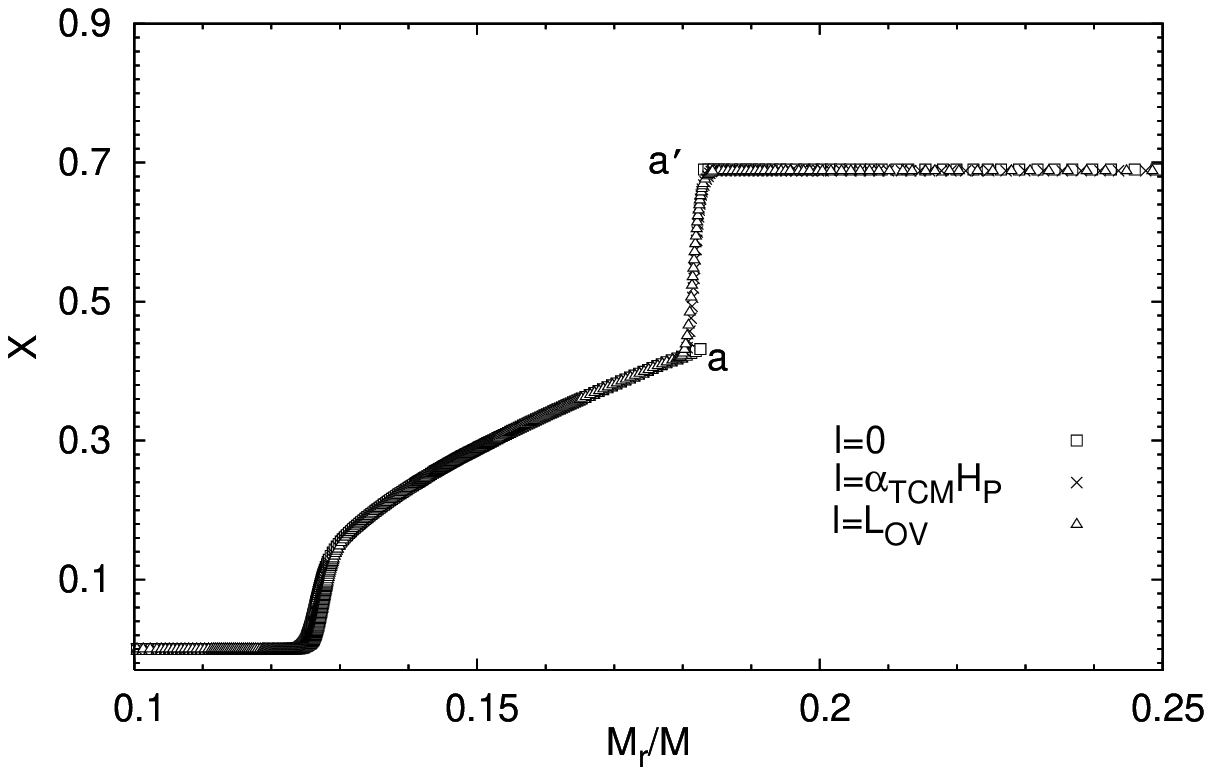}

   \caption{ Hydrogen profile for the star of $5M_{\odot}$ at the phase of the envelope convection penetrating mostly (near the tip of the RGB phase). The case of $l=0$ represents the result obtained from the classical model with the MLT. And $l=\alpha_{TCM}H_{P},L_{OV}$ (two choices of the typical length of the turbulent dissipation stated in Section 2) represent the results from the diffusive mixing model with $C_{X}=10^{-6}$ and $\alpha_{TCM}=0.2$. }
   \label{Fig1}
   \par\vspace{10pt}
   \end{minipage}%
    \begin{minipage}[t]{73mm}
    \centering
   \includegraphics[width=7.2cm, angle=0]{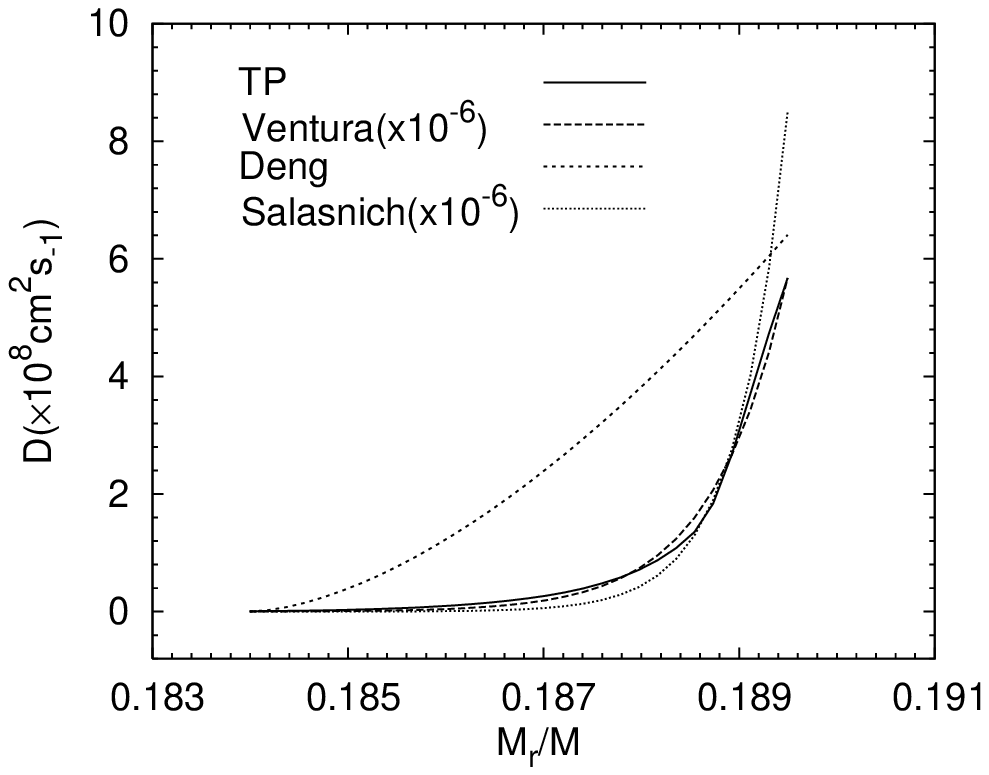}
   \caption{ Comparison of the diffusion coefficient $D$ in the overshooting region below the convective envelope based on a stellar model during the RGB phase for
   the $5M_{\odot}$ star (according to the results from our diffusive mixing model with
   $C_{X}=10^{-6}$ and $\alpha_{TCM}=0.2$). 'TP' represents the result according to the formula in Eq.(2) from this paper, 'Ventura' from Ventura \& D'Antona (2005) (in which $\zeta\cdot f_{thick}=0.03$), 'Deng' from Deng et al. (1996a,b) (in which $P_{dif}=15.0$), and 'Salasnich' from Salasnich et al. (1999) (in which $\alpha_{1}=0.02$).}

   \label{Fig2}
\end{minipage}
   \end{figure}

\subsection{Chemical Profiles in the Stellar Interiors}
The main improvement achieved by our diffusive mixing model is the
modification of the element profiles in the stellar models as shown
Fig.1. In this figure the case of $l=0$ represents the result of the
classical model with the MLT, in which an obvious element abundance
discontinuity (a'-a) can be found around $M_{r}/M=0.18$. This
discontinuity is a direct result of the MLT's assumption of complete
mixing, which can be verified from Fig.4 in the work of Huang \&
Weigert (1983) and Fig.2 in the paper of Stothers \& Chin (1991).
Evidently, it will be removed by using a diffusive mixing model and
replaced by a smooth one, which is behaved by the results
corresponding to either the case of $l=\alpha_{TCM}H_{P}$ or that of
$l=L_{OV}$ in Fig.1. It can be found further that the two different
formulae of the mixing length $l$ result in little difference to the
obtained results. The reason seems that the dimension of the
overshooting region $L_{OV}$ is directly determined by the way of
convective velocity decaying in that region, which is in turn just
determined by $\alpha_{TCM}H_{P}$ appeared in the diffusion terms of
the TCM's equations. However, $l=L_{OV}$ is more physical for it
characterizes how far a convective element can be diffused out from
the edge of the convective unstable zone. The most important is that
there is no additional parameters in this kind of choice, therefore
$l=L_{OV}$ will be adopted in practice for the following
discussions.

In Fig.2 the results of the diffusion coefficient $D$ adopted by
Ventura \& D'Antona (2005), Deng et al. (1996a,b), and Salasnich et
al. (1999) are compared with ours for a model of $5M_{\odot}$ during
the RGB phase. It can be seen that the value of our diffusion
coefficient $D$ is on the whole of the same order of magnitude as
that of Deng et al. (1996a,b), but about six order of magnitude
smaller than those of Ventura \& D'Antona (2005) and Salasnich et
al. (1999) because of the adopted value of $C_{X}$. On the other
hand, however, the e-folding distance of our $D$ is almost the same
as those from Ventura \& D'Antona (2005) and Salasnich et al.
(1999), which are much shorter than that of Deng et al. (1996a,b)
regardless of the value of the parameter $P_{dif}$ in their formula.
Such differences in $D$ lead to different hydrogen profiles in the
overshooting region at the pahse, shown in Fig.3, when the envelope
convection zone penetrates inward mostly.

   \begin{figure}[h!!!]

    \begin{minipage}[t]{71mm}
    \centering

   \includegraphics[width=7.5cm,height=62mm,angle=0]{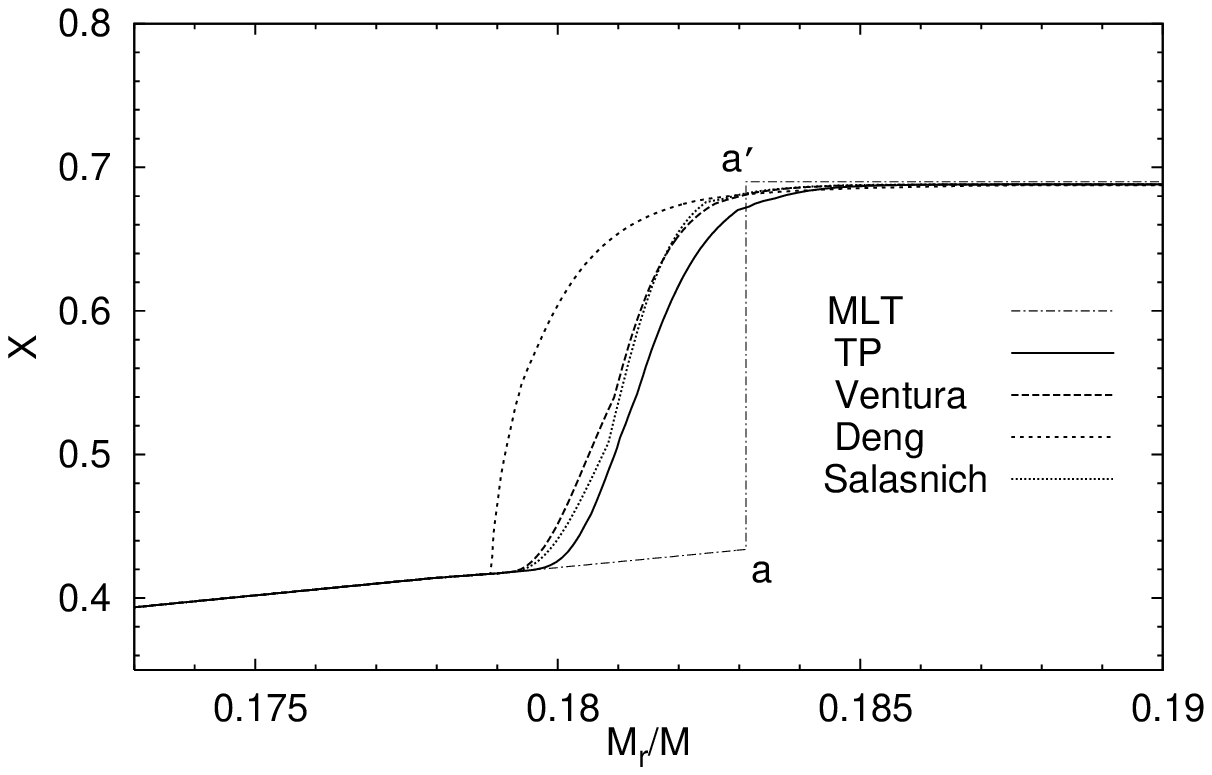}

   \caption{ Hydrogen profile in the star of $5M_{\odot}$ at the phase of the envelope convection penetrating mostly (the same as in Fig.1) based on the different formulae of the diffusion coefficient $D$ as mentioned in Fig.2. 'MLT' represents the result from the classical model with the MLT. }
   \label{Fig3}
   \par\vspace{10pt}
   \end{minipage}%
    \begin{minipage}[t]{71mm}
    \centering
   \includegraphics[width=7.1cm, angle=0]{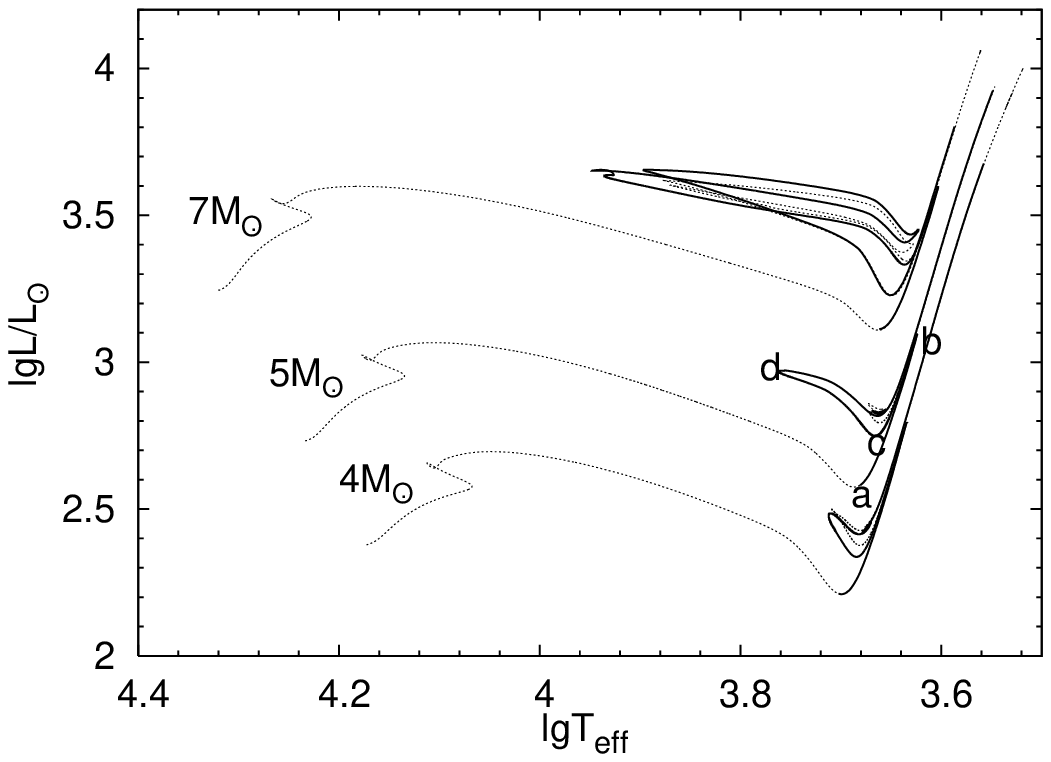}
   \caption{ The HR diagram resulted from the diffusive mixing model (solid lines) and the classical model (dotted lines) for the stars of $4$, $5$ and $7M_{\odot}$. The start of all these solid tracks is at the bottom of the RGB phase up to the AGB phase with $\alpha_{TCM}=0.9$ and $C_{X}=10^{-6}$. The labels (a,b,c,d) beside the curve of the stellar model of $5M_{\odot}$ mark the key positions in the HR diagram, which are described in the text.}

   \label{Fig4}
\end{minipage}
   \end{figure}

\subsection{HR diagram}

Evolutionary tracks based on the diffusive mixing model with those
based on the classical model with the MLT are shown in the HR
diagram in Fig.4. For the sake of convenience of discussions in the
latter section, four letters are pasted at some certain key places
corresponding to: 'a' the bottom of the RGB phase, 'b' the tip of
the RGB phase, 'c' the minimum luminosity descended from the RGB
tip, 'd' the blue endpoint of the blue loop. It is seen in Fig.4
that for the same set of diffusive mixing model's parameters
$\alpha_{TCM}=0.9$ and $C_{X}=10^{-6}$, the lengths of the blue
loops are evidently increased for the models of 5 and 7$M_{\odot}$
and its increment for the model of $5M_{\odot}$ is more prominent
than that of $7M_{\odot}$, but the increment of $4M_{\odot}$ model
is rather small. Such difference found by us are very similar to
those of Alongi et al. (1991) who also focus on the effect of the
envelope overshooting on the extension of the blue loop but with
different mixing scheme. We will concentrate on the results of the
$5M_{\odot}$ model in the following discussions for the mixing
model's parameters $C_{X}$ and $\alpha_{TCM}$ resulting in more
obvious effects on the star of $5M_{\odot}$.

\subsection{Effects of Mixing Model's Parameters $C_{X}$ and $\alpha_{TCM}$}
Results showing the influence of parameter $C_{X}$ on the extension
of the blue loops are given in Fig.5, evolutionary tracks being
plotted from the bottom (except the track of the classical model) of
the RGB phase only to the bluest points of the blue loops for the
sake of clearness. It can be found clearly that the extension of the
blue loops increases with the value of parameter $C_{X}$ when the
value of $\alpha_{TCM}$ is fixed. In turn when the value of $C_{X}$
is fixed the extension of the blue loops increases with the value of
$\alpha_{TCM}$ as shown in Fig.6. The reason for these, will be
analyzed in the following section, is due to the diffusion
coefficient $D$, which is proportional to both of them. From the
right panels of Fig.5 it can be found that the larger the value of
$C_{X}$ is, the lower the RGB tip is obtained in luminosity, when
$\alpha_{TCM}$ is fixed. This conclusion is the similar for the
parameter $\alpha_{TCM}$ as seen in Fig.6, namely the larger the
value of $\alpha_{TCM}$ is, the fainter the RGB tip will be.

   \begin{figure}[h!!!]

   \includegraphics[width=7.0cm,height=10.5cm]{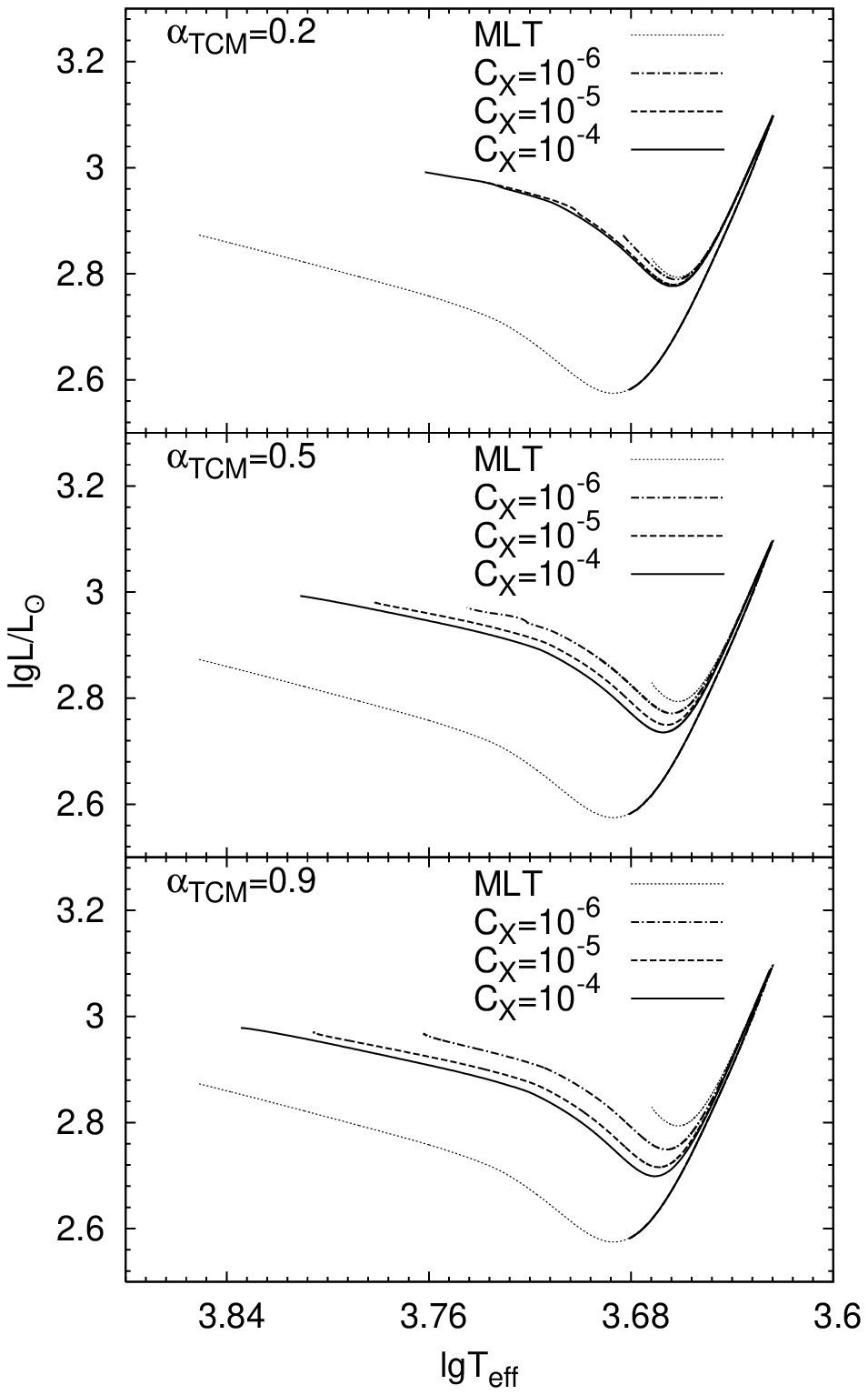}
   \includegraphics[width=11.8cm,height=6.6cm]{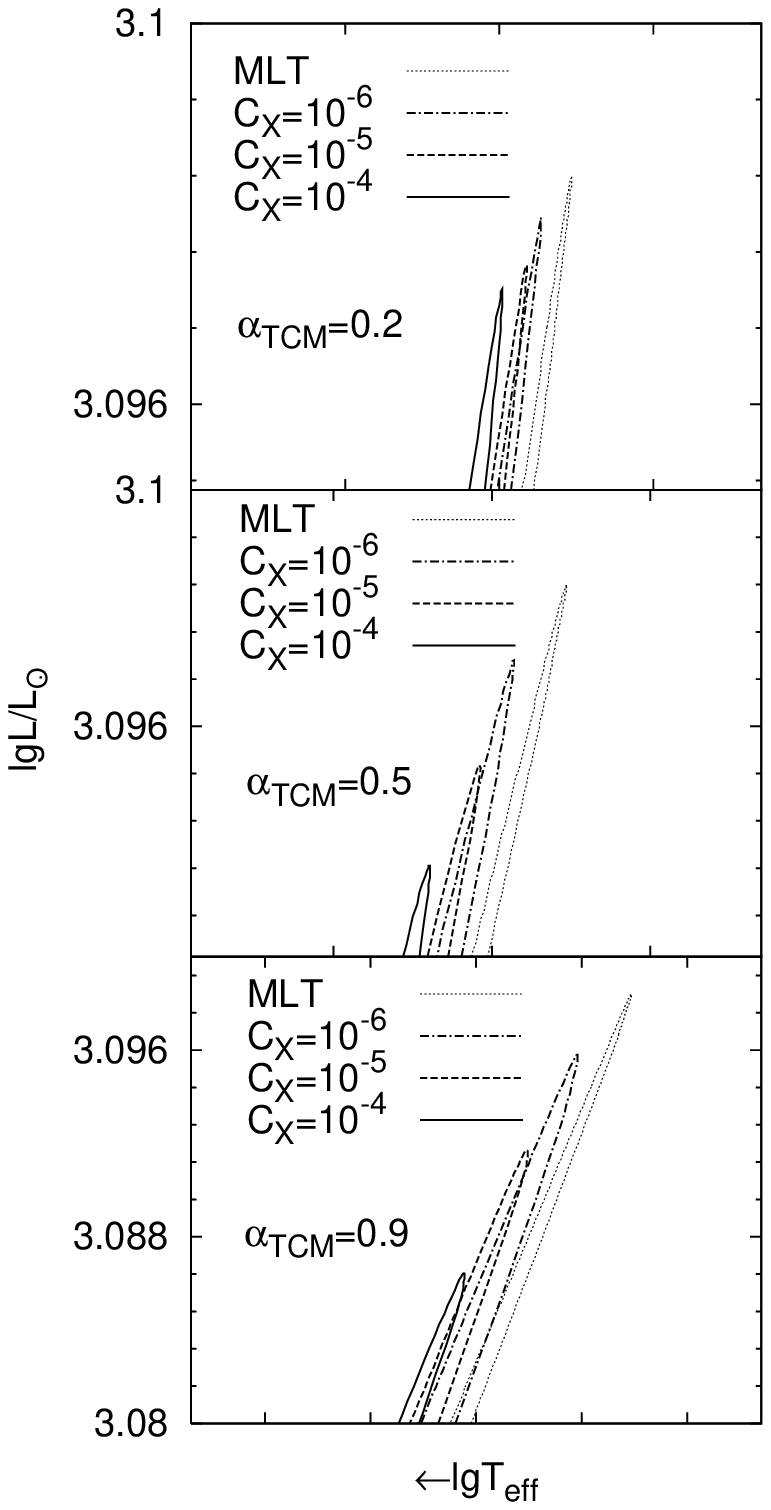}

   \

   \flushleft
   \caption{The left three panels show evolutionary tracks for the star
   of $5M_{\odot}$ from the base of the RGB phase to the bluest points
    (namely from a to d in Fig.4) changing with different
     values of $C_{X}$ when a value of $\alpha_{TCM}$ is fixed.
      The right three panels are the expanded counterparts near
      the tip of the RGB phase in parallel with the left three panels respectively. 'MLT' represents the result
       from the classical model with the MLT.}

   \label{Fig5}
   \end{figure}

   \begin{figure}[h!!!]

   \includegraphics[width=7.0cm,height=10.5cm]{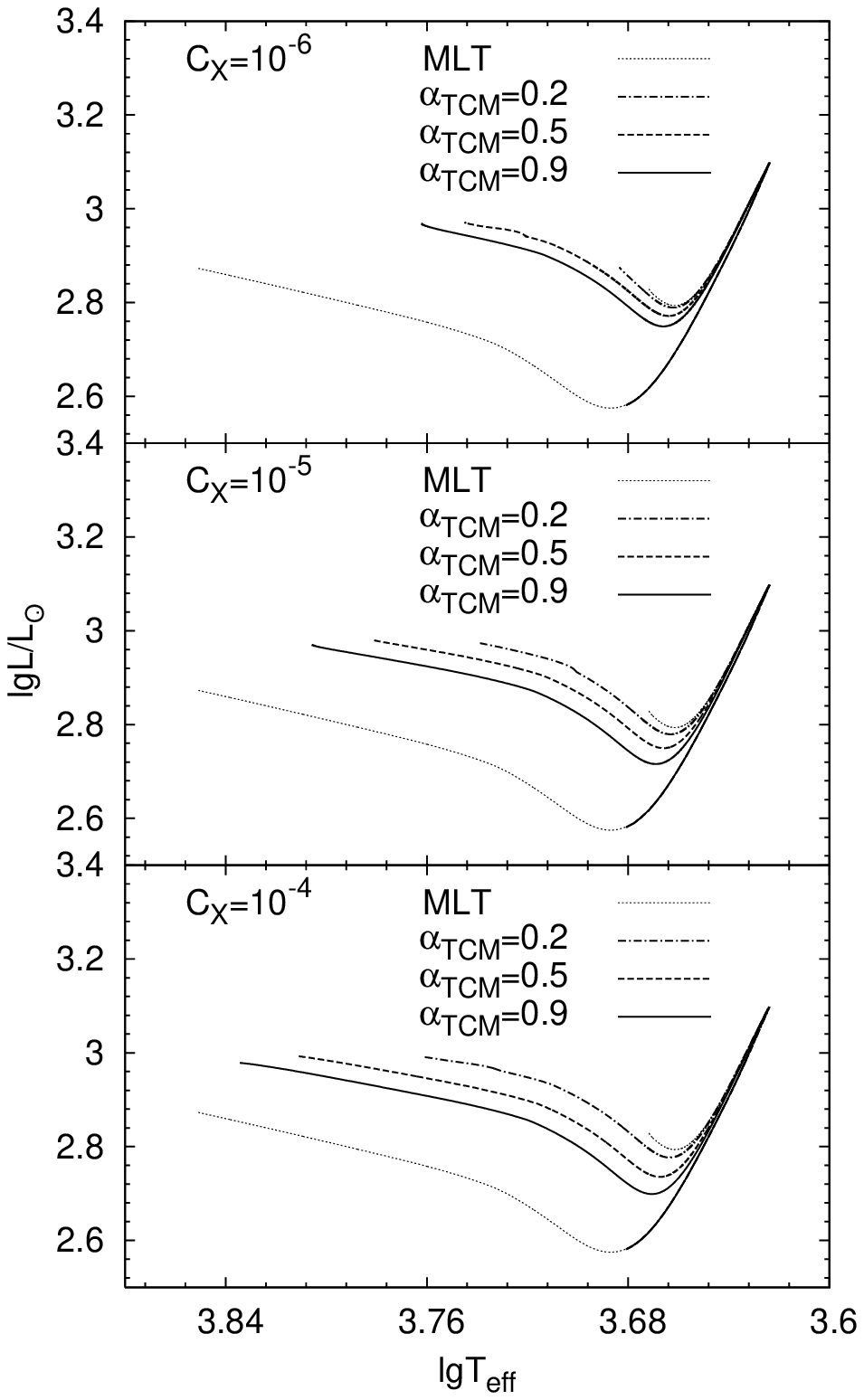}
   \includegraphics[width=11.8cm,height=6.6cm]{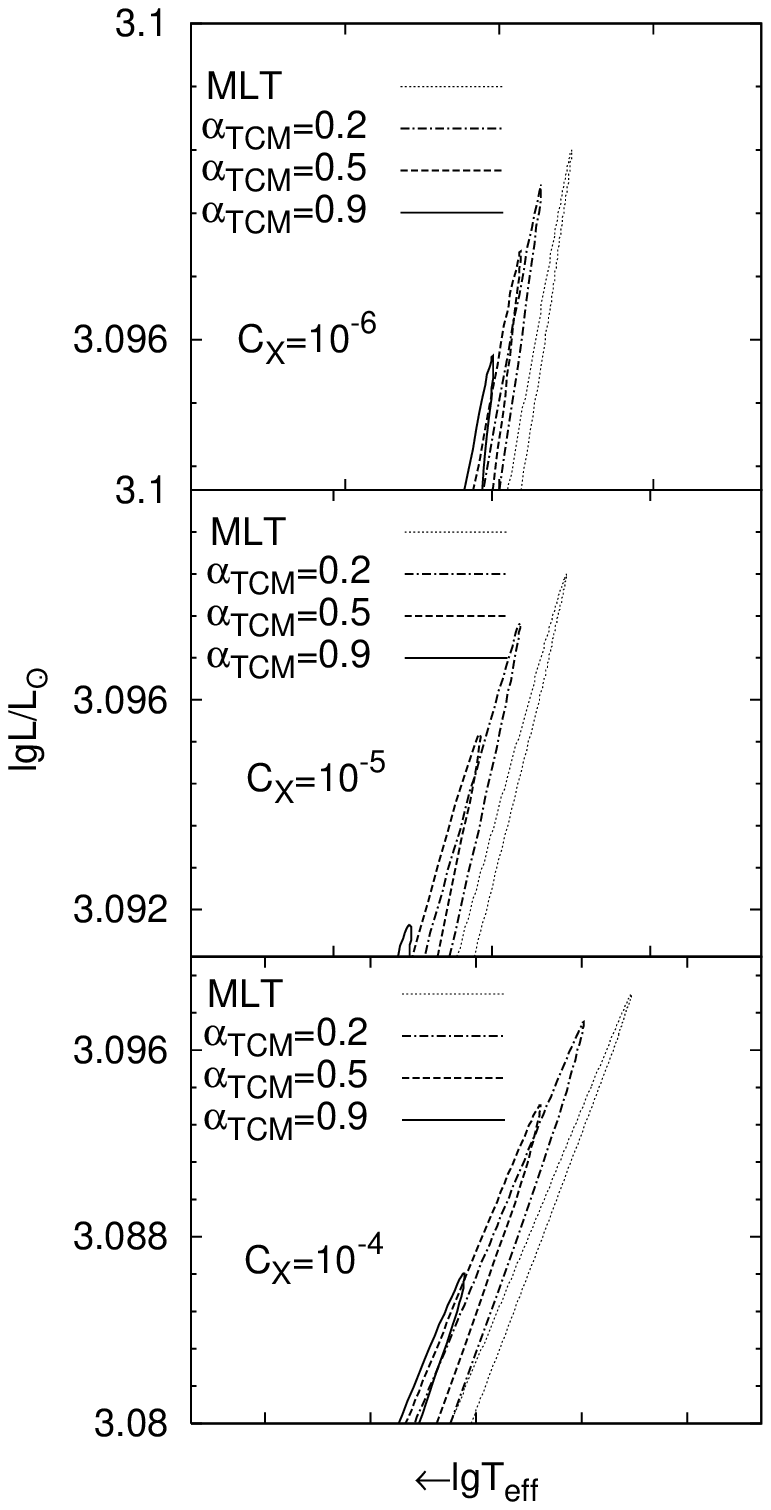}

   \

   \caption{The same as Fig.5, but with different values of  $\alpha_{TCM}$ and a fixed value of $C_{X}$}
   \label{Fig6}
   \end{figure}

\begin{table}

\bc

\begin{minipage}[]{100mm}

\caption[]{ The further increased extent of the dimension of the
convective envelope compared to that of the classical model with the
MLT for the $5M_{\odot}$ star at the phase same as in Fig.7
(approximated by the comparison between the places marked by the
solid star symbols in Fig.7).}\end{minipage}

\small
 \begin{tabular}{ccccccccccc}
  \hline\noalign{\smallskip}
  & $C_{X}=10^{-6}$  &  $C_{X}=10^{-5}$   &    $C_{X}=10^{-4}$\\
  \hline\noalign{\smallskip}
$\alpha_{TCM}=0.2$  & $0.22\%$ & $0.53\%$ &   $0.67\%$ \\
$\alpha_{TCM}=0.5$   & $0.77\%$ &  $1.35\%$ &  $1.86\%$ \\
$\alpha_{TCM}=0.9$   & $1.37\%$ & $2.39\%$ &  $2.90\%$ \\
  \noalign{\smallskip}\hline
\end{tabular}
\ec

\end{table}

\section{discussion}
\subsection {Stellar Structure During the RGB Stage}

As shown in Fig.1 for an example, the element profiles in stars are
modified by the diffusive mixing model. The hydrogen profile based
on different values of $C_{X}$ and $\alpha_{TCM}$ are presented in
Fig.7, all of which belong to evolutionary models around the tip of
the RGB phase (point b). It can be found that: 1) the base of
homogeneous envelops are more extended into stellar interior for all
models with the diffusive mixing model than that of the classical
model withe the MLT does, which are clearly shown by the solid star
symbol in Fig.7 defined by where the hydrogen abundance $X=0.55$.
And the dimension of the convective envelope increases with the
values of $C_{X}$ and $\alpha_{TCM}$. The degree of the envelope
penetration depending on the parameters $C_{X}$ and $\alpha_{TCM}$
is given in Table 1; 2) the structure of the hydrogen profile in the
overshooting region is more or less related to both $C_{X}$ and
$\alpha_{TCM}$. It can be found that the slope of the hydrogen
profile is mainly determined by parameter $C_{X}$, and the hydrogen
profile becomes steeper and steeper when $C_{X}$ increases from
$10^{-6}$ to $10^{-4}$ as seen in either the right or left panels in
Fig.7. On the other hand, however, the overshooting distance is
mainly determined by parameter $\alpha_{TCM}$, and the overshooting
region approaches deeper as the value of $\alpha_{TCM}$ increase.

As convective elements go down into the overshooting region, the
hydrogen abundance will be increased. Such an effect will lead to
the increase of opacity in the overshooting region and thus the
increase of the radiative temperature gradient $\nabla_{rad}$, and
then a convective stable region just below the base of the
convective unstable zone will be converted into a convective
unstable one. As a result the convective envelope will gradually
swallow up its nearby region originally belonging to the
overshooting region along the RGB evolution, resulting in an
appreciable increase of the dimension of the convective envelope. A
direct result of this expansion of convection in the stellar
envelope is the increase of convective heat transfer efficiency,
leading to a smaller stellar radius and a higher effective
temperature seen in Figs.5 and 6. The effect of parameter
$\alpha_{TCM}$ on the stellar chemical structure, at the first
glance, seems confusing for a larger value of $\alpha_{TCM}$ will
normally lead to a smaller convective envelope because of a higher
efficiency of energy transport related to it, which is also
indicated in Fig.4 of Huang \& Weigert (1983). Nevertheless, our
results can be understood by taking the following two reasons into
account. The first one is that we do not take into account the
correction of the temperature structure in the convective envelope
by the TCM, thus the value of parameter $\alpha_{TCM}$ has no effect
on the size of the convective envelope. The second one is related to
the diffusion coefficient $D$ in Eq.(2), in which the effective
mixing length $l$ is chosen to be the dimension of the overshooting
region. Accordingly a larger value of $\alpha_{TCM}$ will lead to a
larger velocity of a convective element at the base of the
convective unstable zone and further lead to a larger distance of
convective overshooting.

Nuclear energy of the hydrogen shell burning resulted from the
diffusive mixing models has been found to be lowered during the RGB
evolutionary stage, which is shown in Fig.8. It can be seen that the
nuclear energy productivity of the H-burning shell is suppressed
more with a larger values of $C_{X}$ and $\alpha_{TCM}$, while that
of the core He-burning is essentially unaffected by the diffusive
mixing of the convective envelope. This is because that the
diffusive mixing in the overshooting region will result in increase
of opacity above the hydrogen burning shell. As a result, some of
heat will be blocked and do mechanical work on this region to make
it expanded a little, resulting in decreases of temperature and
density and hence the H shell nuclear energy generation rate. This
effect explains the depression of luminosity at the tip of RGB phase
(point b) seen in Figs.5 and 6.

   \begin{figure}[h!!!]

   \includegraphics[width=7.1cm]{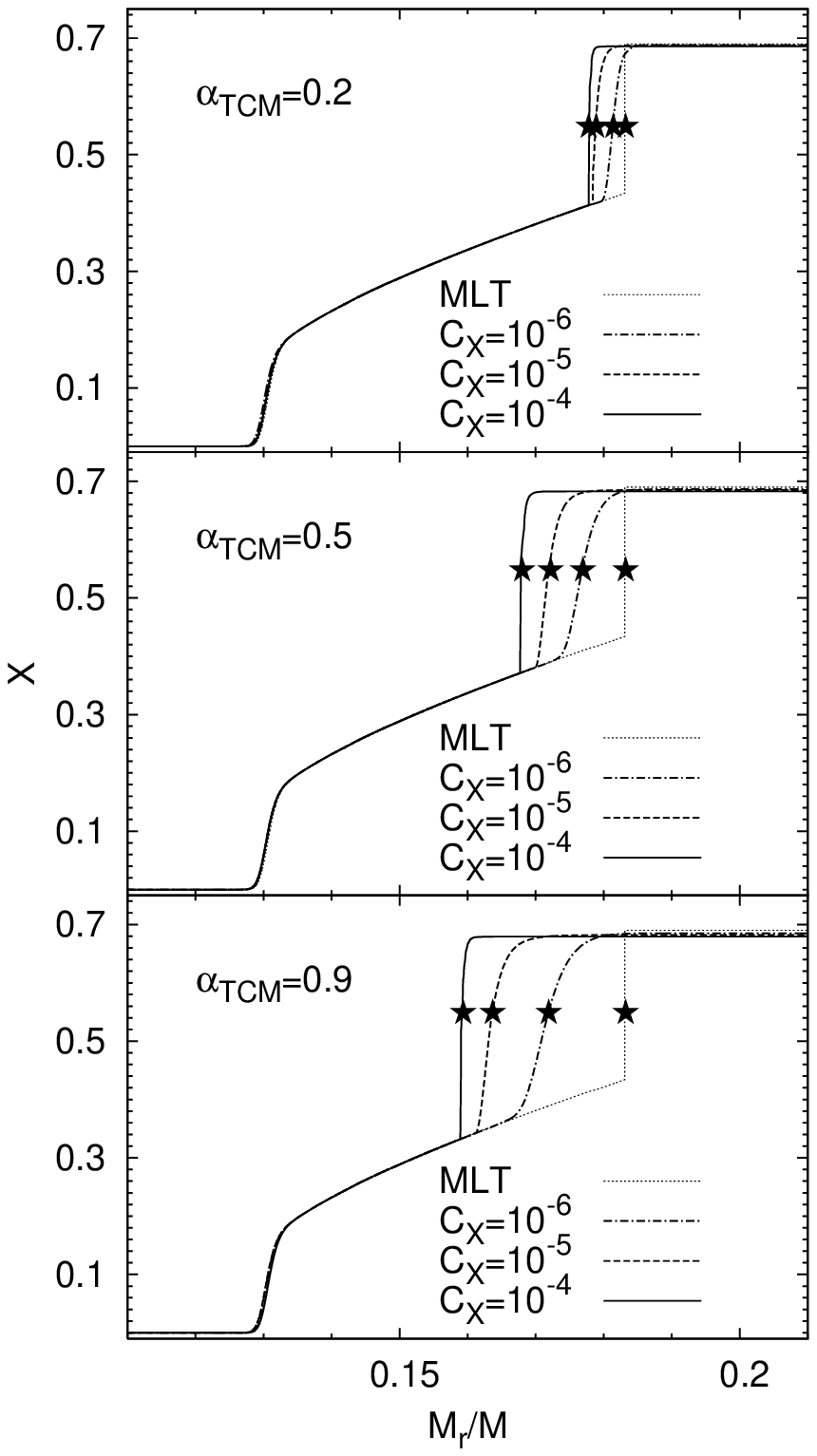}
   \includegraphics[width=7.1cm]{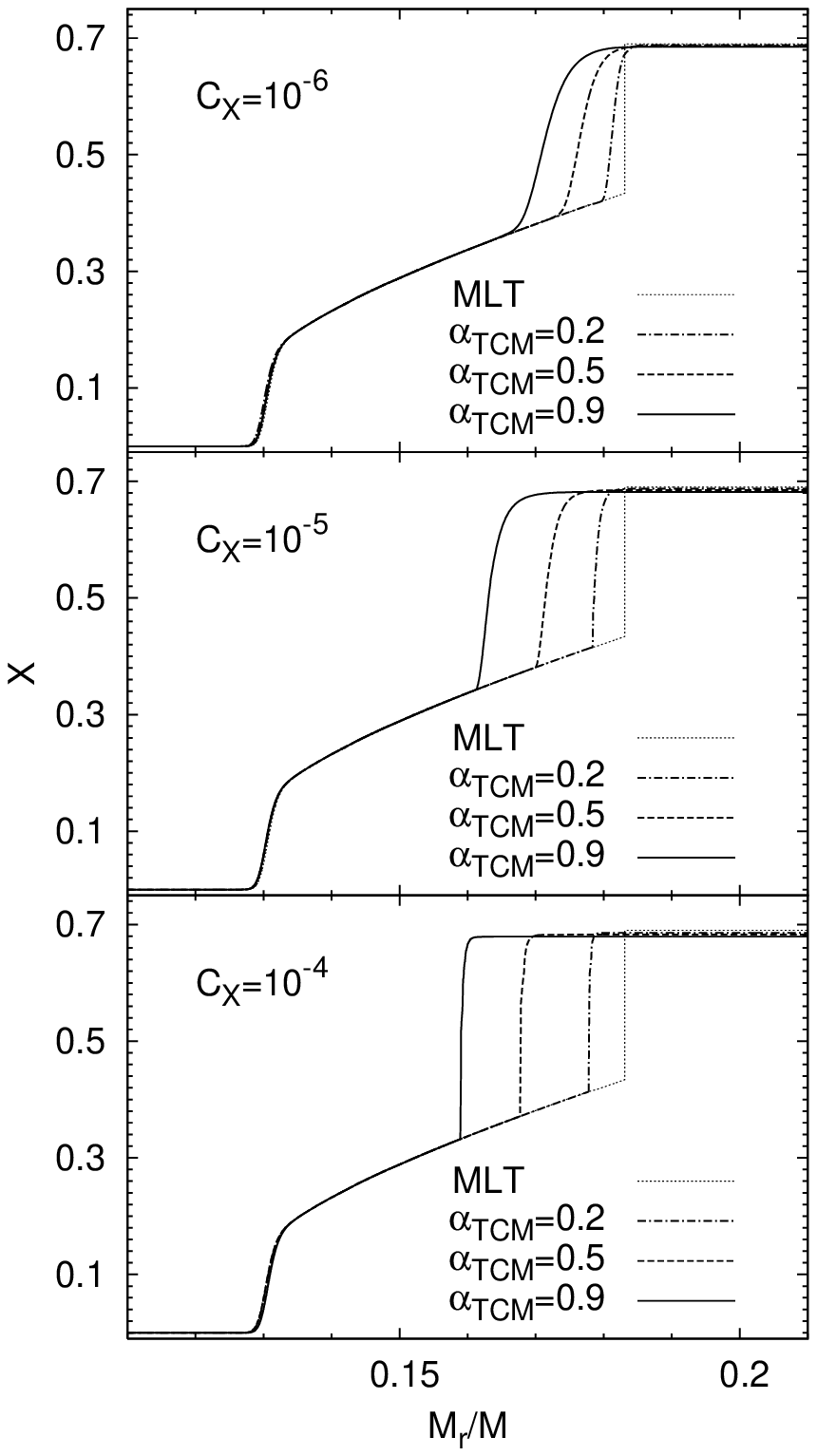}

   \

   \caption{Hydrogen profiles in the star of $5M_{\odot}$ at the phase same as in Fig.1 with different values of $C_{X}$ when a value of $\alpha_{TCM}$ is fixed (left panels) or with different values of $\alpha_{TCM}$ when a value of $C_{X}$ is fixed (right panels). The solid star symbol indicate the locations in stellar interior at hydrogen abundance $X=0.55$ for different sets of the two parameters. 'MLT' represents the result from the classical model with the MLT.}

   \label{Fig7}
   \end{figure}

\subsection{Evolution from the RGB Tip to the Bluest Point}

The most significant effect of the diffusive mixing model on the
evolution of intermediate-mass stars is the development of blue
loop. For the sake of simplicity we only choose the diffusive mixing
model with $C_{X}=10^{-4}$ and $\alpha_{TCM}=0.9$ to make a
comparison with the classical model with the MLT. The obtained
conclusions are also applicable to other combinations of $C_{X}$ and
$\alpha_{TCM}$. During this evolutionary stage the nuclear
luminosity as a function of the central He abundance $(Y_{c})$ are
given in Fig.9. It can be found that as the model evolves forward
beyond the point c the H-shell nuclear luminosity $L_{H}$ and thus
the surface luminosity $L_{S}$ of the diffusive mixing model
increase promptly and ultimately reach a much larger value when the
model arrives at the endpoint of the blue loop compared with the
results of the classical model with the MLT. Before arriving at
point c the luminosities $L_{H}$ and $L_{S}$ of the diffusive mixing
model quickly decrease to their minimum values labeled as c in
Fig.9, which are much lower than the results of the classical model
with the MLT.

   \begin{figure}[h!!!]

    \begin{minipage}[t]{71mm}
    \centering

   \includegraphics[width=7.0cm,angle=0]{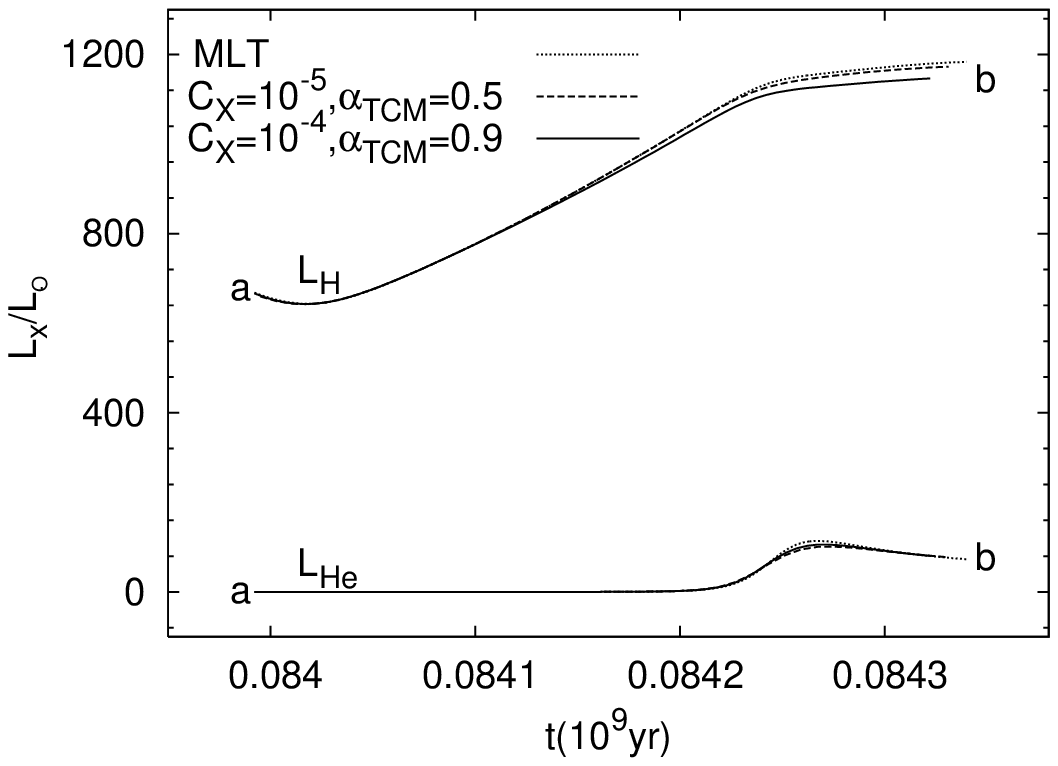}

   \caption{ Nuclear luminosities for the $5M_{\odot}$ star as a function of the age from the point a to point b labelled in Fig.4 with different combinations of parameters $C_{X}$ and $\alpha_{TCM}$, 'MLT' represents the result from the classical model with the MLT. The above three lines labelled as one sign $L_{H}$ are hydrogen shell burning luminosities, the below three lines are central Helium burning luminosities $L_{He}$.}
   \label{Fig8}
   \par\vspace{0pt}
    \end{minipage}
    \begin{minipage}[t]{71mm}
    \centering
   \includegraphics[width=7.0cm, angle=0]{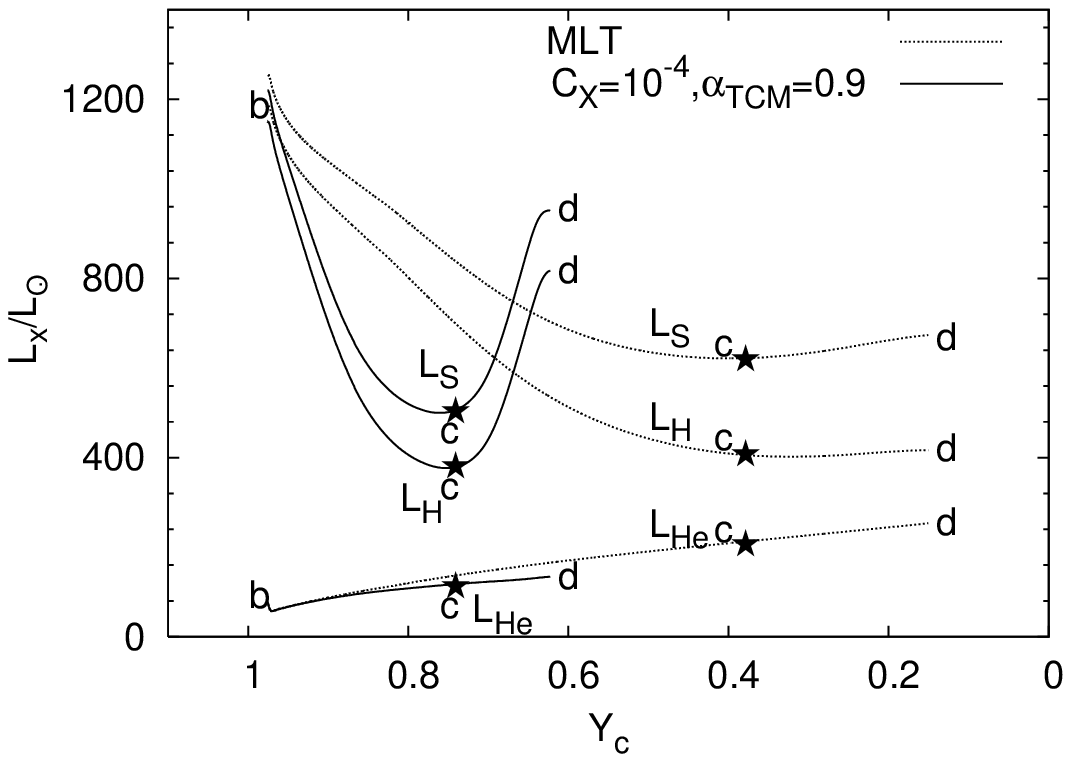}
   \caption{The nuclear luminosities ($L_{H}$ and $L_{He}$) and the surface luminosity ($L_{S}$) for the $5M_{\odot}$ star as a function of the stellar central helium abundance $(Y_{c})$ derived from the classical model with the MLT (dotted lines) and the diffusive mixing model (solid lines) with $C_{X}=10^{-4}$ and $\alpha_{TCM}=0.9$, respectively, which correspond to the phase from the RGB tip to the bluest point in the HR diagram, namely from the point b to point d in Fig.4, and the solid star symbol mark the position of the point c same as in Fig.4.}

   \label{Fig9}
   \end{minipage}
   \end{figure}

As the stellar models evolve from the RGB base (point a) upward to
the RGB tip (point b) the opacity in the overshooting region just
above the H-burning shell will be increased for the diffusive mixing
models. When evolving from the RGB tip to point c such effect will
become more significant, because the H-shell will be more and more
close to the chemical gradient region. In such a way a quick
decrease of the H-shell nuclear luminosity $L_{H}$ will be obtained
for the diffusive mixing model, and thus their convection envelopes
will be receded outward in a faster way during this evolutionary
stage. In addition, the opacity of the whole convective envelope
will be decreased to some extent due to the decrease of hydrogen
abundance in it. Therefore, a much weaker and smaller convection
envelope at point c will be formed for the stars with the diffusive
mixing model.

Some characteristics of the stellar envelope at the bottom of the
RGB (namely point c in Fig.9) are believed to trigger a blue loop.
For example, Renzini et al. (1992) found that if the stellar
luminosity is lower than a critical value $L_{loop}$ at this phase a
blue loop will be developed. However, such criterion is very
sensitive to stellar mass. Afterwards a new criterion, which has
less dependence on the stellar mass, was proposed by Xu \& Li
(2004a). They defined an envelope convection ratio $\eta$ as
$\eta=M_{con}/M_{env}$, where $M_{con}$ is the mass of the
convection zone within the envelope and $M_{env}$ the mass of the
whole envelope above the hydrogen burning shell. Then found if the
value of $\eta$ is smaller than a critical value $\eta_{crit}$, the
envelope will be radiation-dominated and the star will develops a
blue loop. Inversely if the value of $\eta$ is larger than the
critical one the star will has a convection-dominated envelope and
evolve redward after point c in Fig.9. They found that the critical
value $\eta_{crit}$ is to be between 0.24 and 0.37 for a
$5M_{\odot}$ star. For the diffusive mixing model we obtain its
value changing with parameters $C_{X}$ and $\alpha_{TCM}$ given in
Table 2, which is very close to the result of Xu \& Li (2004a). It
can be found that all the values of $\eta$ are smaller than the
classical one, and larger values of parameters $C_{X}$ and
$\alpha_{TCM}$ correspond to smaller $\eta$. Certainly the smaller
the value of $\eta$ is, the convection zone are less extended in the
stellar envelope. As a result a smaller value of $\eta$ corresponds
to a longer blue loop. For example, in Fig.7 for the diffusive
mixing model with $C_{X}=10^{-4}$ and $\alpha_{TCM}=0.9$ the
overshooting region can reach mostly to $M_{r}/M\approx0.16$, which
is obviously more nearer to the star center than
$M_{r}/M\approx0.18$ of the diffusive mixing model with
$C_{X}=10^{-4}$ and $\alpha_{TCM}=0.2$. Correspondingly,
$\eta=0.211$ for $C_{X}=10^{-4}$ and $\alpha_{TCM}=0.9$ is smaller
than $\eta=0.251$ for $C_{X}=10^{-4}$ and $\alpha_{TCM}=0.2$.
Therefore, we can see in Fig.5 or Fig.6 that the former star
develops a longer blue loop than the latter one.

\begin{table}

\bc

\begin{minipage}[]{100mm}

\caption[]{ The value of $\eta$ at point c in Fig.4 varies with
parameters $C_{X}$ and $\alpha_{TCM}$ for the $5M_{\odot}$ star with
the diffusive mixing model, and its value is 0.270 for the classical
model with the MLT.}\end{minipage}

\small
 \begin{tabular}{ccccccccccc}
  \hline\noalign{\smallskip}
 & $C_{X}=10^{-6}$  &  $C_{X}=10^{-5}$   &    $C_{X}=10^{-4}$\\
  \hline\noalign{\smallskip}
$\alpha_{TCM}=0.2$  & $0.258$ & $0.254$ &   $0.251$ \\
$\alpha_{TCM}=0.5$   & $0.249$ &  $0.237$ &  $0.232$ \\
$\alpha_{TCM}=0.9$   & $0.239$ & $0.227$ &  $0.211$ \\
  \noalign{\smallskip}\hline
\end{tabular}
\ec

\end{table}

\section{Comparisons with observations}
\subsection{Instability Strip of the Cepheids}

   \begin{figure}[h!!!]

    \begin{minipage}[t]{71mm}
    \centering

   \includegraphics[width=7.0cm,angle=0]{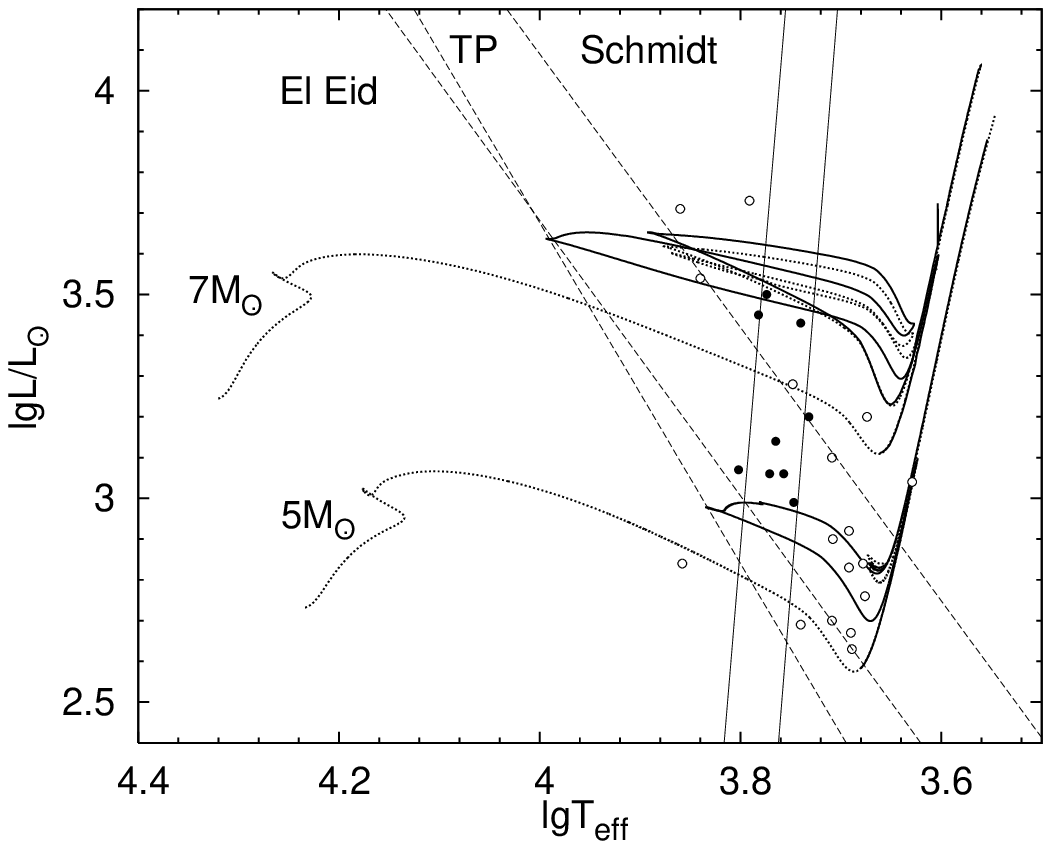}

   \caption{The HR diagram of stellar models of $5$ and $7M_{\odot}$ for the diffusive mixing model with $C_{X}=10^{-4}$ and $\alpha_{TCM}=0.9$ (solid lines) and the classical model with the MLT (dotted lines) comparing to the observed yellow giants and supergiants in open clusters (Schmidt 1984). The solid and open circles respectively denote non-variable and Cephied stars. The blue and red edges of the Cepheid instability strip are from the theoretical result of Li (1993). The three dashed lines are from connecting the two bluest endpoints of the blue loops of the models of $5$ and $7M_{\odot}$, which are labelled as 'El Eid', 'TP' and 'Schmidt' representing respectively the results from the works of El Eid (1995), this paper (TP) by us and Schmidt (1984).}
   \label{Fig10}
   \par\vspace{0pt}
    \end{minipage}
    \begin{minipage}[t]{71mm}
    \centering
   \includegraphics[width=7.0cm, angle=0]{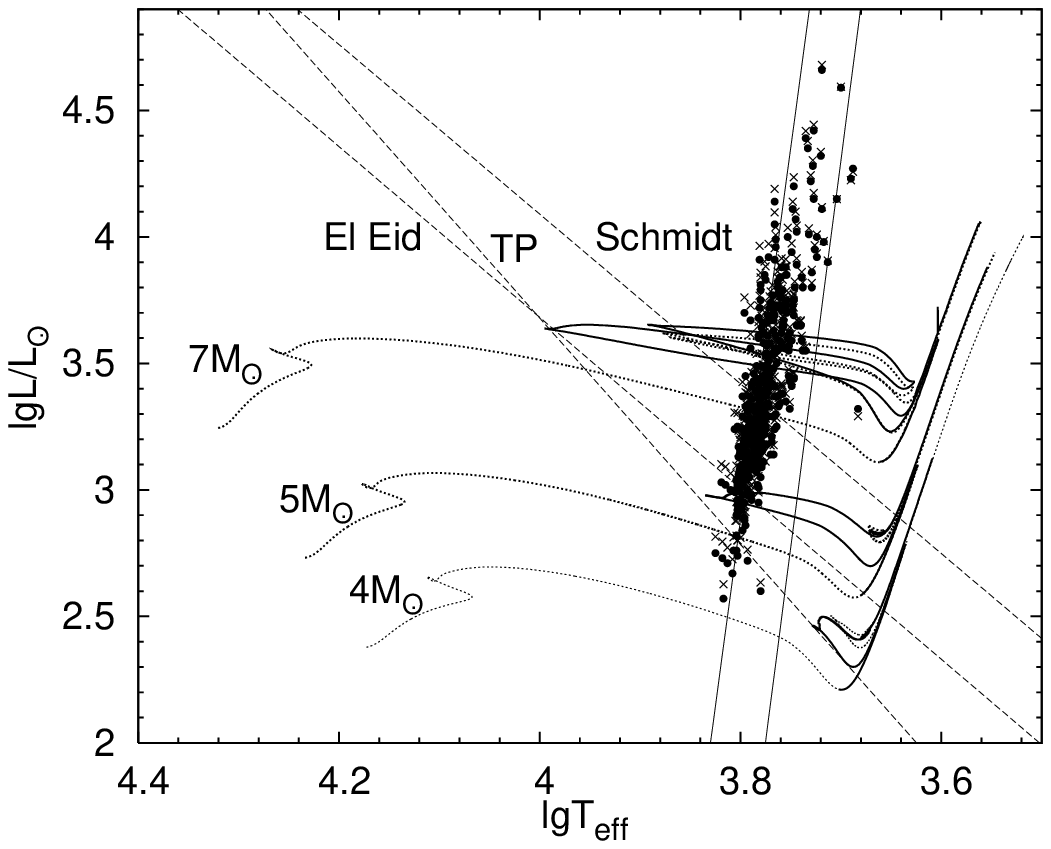}
   \caption{The same as Fig.10 except that the Cepheid stars are replaced by the Galactic classical Cepheids in catalog provided by Fernie et al. (1995). '$\bullet$' represents the result transformed from some photometric data and '$\times$' represents the result from the period-luminosity relation, which are described in detail in the text.}
   \label{Fig11}
   \end{minipage}
   \end{figure}
The observed location of the Cepheid instability strip in the HR
diagram can be used to verify theoretical models and restrict the
models' parameters. In Fig.10, the evolutionary tracks of our $5$
and $7M_{\odot}$ models are compared with 17 non-variable yellow
giants and supergiants and nine Cepheids attaching to the Galactic
open clusters derived from Schmidt (1984) whose theoretical fitting
is also based on the same mass stars with the same initial chemical
content as in the present paper. In order to clearly show the
location of the Cepheids, the theoretical blue and red edges of the
Cepheid instability strip are shown according to the result of Li
(1993). It is obvious that the blue endpoint of the blue loop for
the diffusive mixing model of the $5M_{\odot}$ star can successfully
surpass the predicted blue edge and can also reach the value of the
effective temperature of the bluest observed Cepheid star EV Sct. On
the contrary, the classical model of the same mass only develops a
rather suppressed blue loop and it cannot arrives at even the red
edge of the Cepheid instability strip. However, for the models of
$7M_{\odot}$ both the classical and diffusive mixing model can
successfully predict the existence of the three uppermost Cepheids.
Besides, for the other 17 non-variable stars our diffusive mixing
models of both $5$ and $7M_{\odot}$ are basically in agreement with
the observations. In addition, from Fig.10 we may infer that the
masses of the nine observed Cepheids will be greater than
$5M_{\odot}$ but smaller than $7M_{\odot}$. We use the David Dunlap
Observatory online Galactic classical Cepheids database (Fernie et
al. 1995), for they have small uncertainties in the distance modulus
and color excess (Chiosi et al. 1992). From the database we extract
the mean color excess $E(B-V)$, absolute magnitude $M_{V}$, and
color index $(B-V)$ of all the contained Cepheids (about 500) to get
their distribution in the HR diagram through the following
transforms: 1) the effective temperature: $\lg
T_{eff}=3.886-0.175(B-V)_{0}$, where $(B-V)_{0}$ is the intrinsic
color index (Kraft 1961); 2) the luminosity: $\lg
L/L_{\odot}=(4.75-M_{bol})/2.5$, where $M_{bol}=BC+M_{V}$ is the
absolute bolometric magnitude and $BC=0.15-0.322(B-V)_{0}$ is the
bolometric correction (the same as in Schmidt 1984). The result is
shown in Fig.11 and represented by '$\bullet$' symbol.
Simultaneously the luminosities of these Cepheids are verified by
the period-luminosity relation $\lg L/L_{\odot}=2.43+1.179lgP$
(resulted from the period-radius relation of Gierren et al. (1989),
the relation of the color index and period of Fernie (1990), and the
relation of the color index and effective temperature of Kraft
(1961)) and represented by '$\times$' symbol. We have found that the
difference of the obtained results from the two methods is very
small and the relative difference is less than $3\%$. From Fig.11 it
can be found that the blue loop of the $5M_{\odot}$ star with the
diffusive mixing model is in good agreement with the observations,
its bluest endpoint perfectly arriving at the position of the bluest
Cepheid.

The result of Schmidt (1984) is also shown in Fig.10 and Fig.11,
those stellar models being also based on the same masses of $5$ and
$7M_{\odot}$ with the same chemical compositions (Y, Z)=(0.28,
0.02). It can be seen that the dashed line labeled with 'Schmidt',
which connecting the two bluest endpoints of the blue loops
belonging respectively to the $5$ and $7M_{\odot}$ models, is
located at a too much luminous place, and can not predict those
observed less massive Cepheids below it. El Eid (1995) made some
improvements on the opacities and nuclear reactions into their
evolutionary code, and the element mixing in both core and envelope
convection regions being treated as a diffusive process. The
obtained results are shown as a dashed line labeled with 'El Eid' in
Fig.10 and Fig.11. It can be seen that their results predict most of
the observed low mass Cepheid and are in better agreement with the
observations. The dashed line labeled with 'TP' based on our results
is steeper than the others. In other words, our theoretical results
can predict more Cepheids with smaller masses. But there are still a
few low mass Cepheids shown in Fig.11 below the predicted region.

\subsection{Number Ration $N_{B}/N_{R}$ of Blue to Red Stars}

\begin{table}

\bc

\begin{minipage}[]{100mm}

\caption[]{The $\tau_{He}^{B}/\tau_{He}^{R}$ ratio of the core
helium burning time spent at the blue and red side of the HR diagram
depending on parameters $C_{X}$ and $\alpha_{TCM}$ for the model of
$5M_{\odot}$ with the diffusive mixing model, and its value is 0.0
for the classical model with the MLT.}\end{minipage}

\small
 \begin{tabular}{ccccccccccc}
  \hline\noalign{\smallskip}
 & $C_{X}=10^{-6}$  &  $C_{X}=10^{-5}$   &    $C_{X}=10^{-4}$\\
  \hline\noalign{\smallskip}
$\alpha_{TCM}=0.2$  & $0.000$ & $0.107$ &   $0.141$ \\
$\alpha_{TCM}=0.5$   & $0.183$ &  $0.333$ &  $0.486$ \\
$\alpha_{TCM}=0.9$   & $0.326$ & $0.642$ &  $0.843$ \\
  \noalign{\smallskip}\hline
\end{tabular}
\ec

\end{table}

\begin{table}

\bc

\begin{minipage}[]{100mm}

\caption[]{The same as Table 3 but for the model of $7M_{\odot}$,
and $\tau_{He}^{B}/\tau_{He}^{R}=0.283$ for the classical model with
the MLT.}\end{minipage}

\small
 \begin{tabular}{ccccccccccc}
  \hline\noalign{\smallskip}
 & $C_{X}=10^{-6}$  &  $C_{X}=10^{-5}$   &    $C_{X}=10^{-4}$\\
  \hline\noalign{\smallskip}
$\alpha_{TCM}=0.2$  & $0.309$ & $0.399$ &   $0.485$ \\
$\alpha_{TCM}=0.5$   & $0.598$ &  $0.739$ &  $0.868$ \\
$\alpha_{TCM}=0.9$   & $0.668$ & $1.053$ &  $1.342$ \\
  \noalign{\smallskip}\hline
\end{tabular}
\ec

\end{table}

   \begin{figure}[h!!!]
   \par\vspace{-30pt}
   \centering
   \includegraphics[width=8.0cm,height=6.6cm]{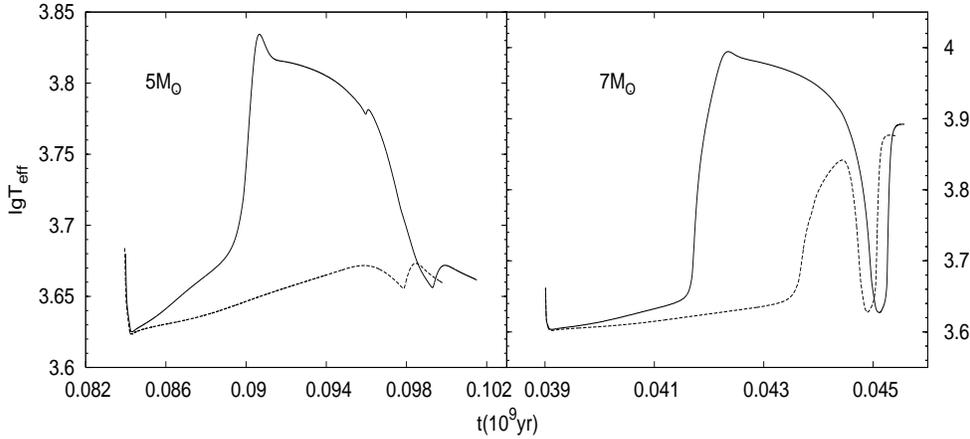}
   \quad
   \quad  \quad
   \quad  \quad
   \quad  \quad
   \quad  \quad
   \quad  \quad
   \caption{Effective temperatures of the $5$ and $7M_{\odot}$ stars as a function of the age (in unit of $10^{9}$ yr) during the core helium burning
   phase. In both panels solid lines stand for the diffusive mixing model with $C_{X}=10^{-4}$ and $\alpha_{TCM}=0.9$ and dashed lines stand for the classical model with the MLT
   .
   }
   \label{Fig12}
   \end{figure}

As we have mentioned in the Introduction, the observed ratio
$N_{B}/N_{R}$ of blue to red giants can be used to probe stellar
models related to some physic processes, for example, the
convection. On the theoretical side, the counterpart of
$N_{B}/N_{R}$ is usually referring to the time ratio of the core
helium burning staying at the blue and red side in the HR diagram,
namely $\tau_{He}^{B}/\tau_{He}^{R}$ , and the dividing line is
taken at $\lg T_{eff}=3.7$ as in the works of Stothers \& Chin
(1993) and El Eid( 1995). The dividing line defined above just
corresponds to the one of the classification of blue and red
supergiants according to the spectral classes, that is, the boundary
between G and K spectral classes. The values of
$\tau_{He}^{B}/\tau_{He}^{R}$ depending on the diffusive mixing
model's parameter $C_{X}$ and the TCM's parameter $\alpha_{TCM}$ are
given in Table 3 and Table 4, respectively, for the $5$ and
$7M_{\odot}$ star. In Table 3 it can be found that the value of
$\tau_{He}^{B}/\tau_{He}^{R}$ increases with each of the two
parameters when the other is fixed, which is in accordance with the
extension of the blue loop depending on the two parameters. It can
be seen in Table 3 that for the diffusive mixing models with the
parameters in the ranges $0.5\leq \alpha_{TCM}\leq0.9$ and
$10^{-5}\leq C_{X}\leq 10^{-4}$, the obtained
$\tau_{He}^{B}/\tau_{He}^{R}$ is confined in the range
0.333$\sim$0.843, which cover the results 0.58 or 0.69 in the work
of El Eid (1995) depending on different convection models, but lower
than 1.152 obtained in the paper of Deng et al. (1996b). Similarly
the value of $\tau_{He}^{B}/\tau_{He}^{R}$ for the star of
$7M_{\odot}$ shown in Table 4 also increases with the values of the
two parameters, and the largest one is equal to 1.342, which is very
close to the values 1.33 in the work of El Eid (1995), 1.66 (in
Maeder \& Meynet 1989), 0.5$\sim$2.1 (in Carson \& Stothers 1976),
1.44 (in Bertelli et al. 1985), and 1.263 (in Deng et al. 1996b).
However, these values are derived from different treatments of
convection. It needs to point out that for the classical model with
the MLT the $\tau_{He}^{B}/\tau_{He}^{R}$ is 0.284 for $7M_{\odot}$
star and zero for $5M_{\odot}$ star. These results manifest that the
diffusive mixing model can significantly extend the duration of the
models staying in the blue side of the HR diagram, even though the
length of the blue loop for the $7M_{\odot}$ star is comparable with
that of the classical model with the MLT as shown in Fig.4. This can
be further confirmed by the result in Fig.12. The reason is that the
blue loop for the diffusive mixing model will be formed at a higher
value of the central He abundance compared with the case of the
classical model with the MLT.

The value of $N_{B}/N_{R}$ for one cluster suffers from many
uncertainties. In general, the incompleteness of the number of the
observed red stars is more severe than that of the blue stars,
because faint and highly reddened stars have simply not been
observed (Humphreys \& McElroy 1984). This will lead to a higher
$N_{B}/N_{R}$ than the real one and lead to an inaccurate value. The
photometric data suffer from uncertainties of the distance modulus
and color excesses accounted for the interstellar extinction. They
are differently based on $UBV$, $uvby$, $H\beta$, $H\gamma$
photometry (Stothers 1991). Turner et al. (1998) give the average
uncertainty in the color index $(B-V)$ being $\pm 0.02$. Therefore,
the average uncertainties of the distance modulus and color excesses
are $\pm 0.06$ and $\pm 0.02$, respectively. Furthermore, the
uncertainties in lg$T_{eff}$ and lg$L/L_{\odot}$ are respectively
$\pm0.0035$ and $\pm0.0026$. And the conversion to the luminosities
and effective temperatures of stars is of uncertainty based on the
photometric data. For example, if the conversion of $\lg
T_{eff}=3.886-0.175(B-V)_{0}$ is adopted, the uncertainty of
lg$T_{eff}$ is $\pm0.02$ (Schmidt 1984). However, the spectroscopic
data can provide a more accurate assessment of the effective
temperature, but the star counts available is comparatively small
(Eggenberger et al. 2002). Contamination by field stars can increase
both $N_{B}$ and $N_{R}$. The separation of field star from cluster
members can be done using information of proper motion, radial
velocities, spectral types, color-color diagram and so on (Meylan \&
Maeder 1982). The proper motion is regarded as a better one (Mohan
\& Sagar 1988), but it is hard to obtain its observational data
(Harris 1976). Harris (1976) found that the uncertainty caused by
the field star contamination increase with the age of the open
cluster age group. For his 'III, IV' age groups (the involved
evolving stars in region of $5-8M_{\odot}$) we find that the
uncertainty of $N_{B}/N_{R}$ due to the field star is $\pm0.088$.
With respect to the uncertainty caused by the dynamical evolution of
a cluster, during the last phase of lifetime of a cluster low-mass
stars will be evaporated due to the relaxation process of star-star
encounters, while more massive stars will sink towards the cluster
center (Nordstr\"{o}m et al. 1997; Vesperini et al. 2009). However,
before such phase high-mass stars will evolve for at least $10^{8}$
year (the typical lifetime of open clusters) (Lada \& Lada 2003),
and some blue and red giants will be produced for such time
interval. Most of these giants will not be ejected when the
evaporation of stars occurs, because they are more massive ones.
Therefore, the dynamical evolution of a cluster has less significant
effect on the value of $N_{B}/N_{R}$. Furthermore, the value of
$N_{B}/N_{R}$ has been shown to vary through galaxies and clusters
(e.g. Langer \& Maeder 1995). In addition, based on spectroscopic
data the star counts $N_{B}$ is referred to either O, B, A, F, G
type stars (e.g. Deng et al. 1996b) or only O, B, A type ones (e.g.
Eggenberger et al. 2002). And there is a relative error (no more
than $10\%$) of the ratio $\tau_{He}^{B}/\tau_{He}^{R}$ rested on
whether the breathing convection (Castellani et al. 1985) is
occurred or not. Therefore, taking these uncertainties into account
the average $N_{B}/N_{R}$ of some clusters are usually adopted to
make a comparison with a theoretical counterpart, and we merely
expect the agreement between the value of $N_{B}/N_{R}$ and that of
$\tau_{He}^{B}/\tau_{He}^{R}$ to be achieved to a certain extent.

In the following comparisons only the Galactic open clusters are
concerned, in order to satisfy the requirements of the input physics
of the stellar models described in Section 3. In the work of Carson
\& Stothers (1976) the values of $N_{B}/N_{R}$ were found to be 0.33
and about 1.0 respectively for their middle-age clusters and young
ones, and the correlated initial masses of the evolved giants for
the two age groups they obtained, were roughly equal to $5$ and
$7M_{\odot}$ respectively. It is obvious that the results of our
classical models with the MLT of $5$ and $7M_{\odot}$ can not
achieve these observed values. However, our results with the
diffusive mixing model in Table 3 and Table 4 can cover them. It can
be found that the parameter set of $C_{X}=10^{-5}$ and
$\alpha_{TCM}=0.5$ or $C_{X}=10^{-6}$ and $\alpha_{TCM}=0.9$ for the
star of $5M_{\odot}$ and a particular combination of the model
parameters in the range of $0.5\leq \alpha_{TCM}\leq0.9$ and
$10^{-5}\leq C_{X}\leq 10^{-4}$ for the star of $7M_{\odot}$ result
in the best fit to the observations. There are comparatively more
stars to be counted in the paper of Stothers (1991). The value of
$N_{B}/N_{R}$ in their intermediate age clusters corresponding to
intermediate mass stars in the range $4$ to $9M_{\odot}$ is from
0.37 to 0.55 depending on the adopted sources and the average for
these sources is 0.42. Comparing to our results, we find that our
values of $\tau_{He}^{B}/\tau_{He}^{R}$ for the $5M_{\odot}$ star in
Table 3 can reproduce those values, and the two model parameters can
be confined accordingly in the range of $0.5\leq
\alpha_{TCM}\leq0.9$ and $10^{-5}\leq C_{X}\leq 10^{-4}$.
Nevertheless, for the $7M_{\odot}$ star two suitable ranges are
obtained: $0.2\leq \alpha_{TCM}\leq0.5$ when $C_{X}=10^{-6}$ or
$10^{-5}\leq C_{X}\leq 10^{-4}$ when $\alpha_{TCM}=0.2$.
Fortunately, the correlated initial masses to the source of Harris
(1976) adopted by them is very close to $5M_{\odot}$, and the value
of $N_{B}/N_{R}$ from this source is 0.55 for both the III and IV
age groups. From this value we find that the values of the two model
parameters can still be confined in the same range. With respect to
our model of $7M_{\odot}$, the result of Maeder \& Meynet (1989) is
taken into account, in which $N_{B}/N_{R}$ is in a range from 0.55
to 3.0 and its average value is about 1.0 as in the paper of Carson
\& Stothers (1976). Furthermore, the above average value of Maeder
\& Meynet(1989) should be corrected to be in the range of 1.0 to 1.3
accounting for throwing off one unusual young cluster according to
their own analysis. It can be seen that these observed values are
close to our results of $\tau_{He}^{B}/\tau_{He}^{R}$ for the
$7M_{\odot}$ star with the diffusive mixing model of $10^{-5}\leq
C_{X}\leq 10^{-4}$ when $\alpha_{TCM}=0.9$.

\section{ Concluding remarks}
The principal aim of this work is to study the effect of chemical
mixing by turbulent convection in the convective envelope of
intermediate-mass stars on the formation and extension of the blue
loop in the HR diagram based on the TCM proposed by Li \& Yang
(2007). We have used the TCM to self-consistently obtain the whole
outer convective envelope including the so-called overshooting
region. The chemical mixing in such region is disposed as a
diffusive process, of which the diffusive coefficient completely
depends on the characteristics of the turbulent convection in this
region. We have proposed a new reasonable formula of the diffusion
coefficient. There are two parameters concerned by us. They are
respectively diffusion parameter $C_{X}$ in the diffusive mixing
model and parameter $\alpha_{TCM}$ similar to the mixing length
parameter in the MLT. We have presented evolutionary sequences for
stellar models of $4$, $5$ and $7M_{\odot}$. Different values of the
two parameters at last result in differences among the obtained blue
loops, which critically depends on the correction of the hydrogen
profile above the H burning shell, and thus result in the different
theoretical predictions of the observed number ratio of stars in the
blue and red part of the HR diagram. The conclusions can be
summarized as follows.

\hangafter=1\setlength{\hangindent}{2.8em}1) Comparing with the
results of the classical model with the MLT, the main improvement
achieved by the diffusive mixing model is the correction of the
element profiles above the H burning shell during the RGB phase. Two
choices of the mixing length $l$ appeared in the formula of the
diffusive coefficient in Eq.(2), namely $l=\alpha_{TCM}H_{P}$ and
$l=L_{OV}$, lead to a very similar correction of the element
profiles and thus have little difference in the later evolutionary
results. And the slope of the hydrogen profile in the overshooting
region is mainly determined by parameter $C_{X}$, and the
overshooting distance is mainly determined by parameter
$\alpha_{TCM}$.

\hangafter=1\setlength{\hangindent}{2.8em}2) The length of the blue
loop is increased with different combinations of the values of
$C_{X}$ and $\alpha_{TCM}$ compared to the results of the classical
model with the MLT, and the increment for the model of $5M_{\odot}$
is more prominent than that of the $7M_{\odot}$ star for a same set
of the two parameters, but the increment for the model of
$4M_{\odot}$ is less obvious. In addition, a larger value of $C_{X}$
is given, a lower luminosity of the RGB tip and a hotter effective
temperature of it are obtained, when $\alpha_{TCM}$ is fixed, and
vice versa.

\hangafter=1\setlength{\hangindent}{2.8em}3) Due to the increase of
opacity in the overshooting region caused by the diffusive mixing
model, the H-shell nuclear luminosity $L_{H}$ will be decreased in a
much faster manner when the star evolves from the RGB tip to its
bottom. And thus the outer homogeneous convection zone will be
receded outward in a faster way that there is less energy provided
by the H-shell burning. Additionally, opacity in the homogeneous
convection zone will be decreased. At last a much weaker and smaller
convection zone in the stellar envelope will be formed at the base
of the RGB, which then leads to a longer blue loop.

\hangafter=1\setlength{\hangindent}{2.8em}4) Comparing to the
observed location of the Cepheid instability strip the parameters of
our diffusive mixing model are confined in a range: $0.5\leq
\alpha_{TCM}\leq0.9$ and $10^{-5}\leq C_{X}\leq 10^{-4}$.

\hangafter=1\setlength{\hangindent}{2.8em}5) Our results of
$\tau_{He}^{B}/\tau_{He}^{R}$ based on the diffusive mixing model
are on the whole in accordance with not only other theoretical ones
but also the observations. By comparing our results of the ratio
$\tau_{He}^{B}/\tau_{He}^{R}$ with the observed number ratio
$N_{B}/N_{R}$ of blue to red evolved stars in the Galactic open
clusters we find that the parameters of the diffusive mixing model
for the $5M_{\odot}$ star should be confined to: $0.5\leq
\alpha_{TCM}\leq0.9$ and $10^{-5}\leq C_{X}\leq 10^{-4}$, which is
the same range as the result from the comparisons of the location of
the Cepheid instability strip. While for the $7M_{\odot}$ star, the
observational data with considerable scatters correspond to several
possible combinations of the parameters such as $0.2\leq
\alpha_{TCM}\leq0.5$, and $ C_{X}=10^{-6}$ or $10^{-5}\leq C_{X}\leq
10^{-4}$ when $\alpha_{TCM}=0.2$ or $10^{-5}\leq C_{X}\leq 10^{-4}$
when $\alpha_{TCM}=0.9$. Anyway, our values derived from the
diffusive mixing model within the ranges of $C_{X}$ and
$\alpha_{TCM}$ adopted in the present paper can almost cover these
different observational data.

It needs to point out that there are many factors to affect the
properties of the blue loop, as Iben (1993) have already argued that
the properties of the blue loop cannot be related to a single factor
or effect. Many effects are still not fully understood (El Eid 1995;
Valle et al. 2009), especially for the effect of convection.

\normalem
\begin{acknowledgements}
This work is supported by the National Natural Science Foundation of
China (Grant No. 10973035 and No. 10673030). Fruitful discussions
with Q.-S. Zhang, C.-Y. Ding and J. Su are highly appreciated.
\end{acknowledgements}

\label{lastpage}

\end{document}